\documentclass[10pt,aps,prd,amsmath,amssymb,superscriptaddress,twocolumn,nofootinbib,showpacs,preprintnumbers,amsfonts,notitlepage]{revtex4-1}

\usepackage{preamble}
\usepackage{affiliations}
\usepackage{orcidlink}

\begin{document}
\preprint{JLAB-THY-24-4182}
\title{Towards a unified description of hadron scattering at all energies}

%% AUTHORS
\author{D.~\surname{Stamen}\orcidlink{0000-0003-0036-2928}}
\email{stamen@hiskp.uni-bonn.de}
\affiliation{\HISKP}
\author{D.~\surname{Winney}\orcidlink{0000-0002-8076-243X}}
\email{winney@hiskp.uni-bonn.de}
\affiliation{\HISKP}
\author{A.~\surname{Rodas}\orcidlink{0000-0003-2702-5286}}
\affiliation{\odu}
\affiliation{\jlab}
\author{C.~\surname{Fern\'andez-Ram\'irez}\orcidlink{0000-0001-8979-5660}}
\affiliation{\uned}
\author{V.~\surname{Mathieu}\orcidlink{0000-0003-4955-3311}}
\affiliation{\ub}
\author{G.~\surname{Monta\~na}\orcidlink{0000-0001-8093-6682}}
\affiliation{\jlab}
\author{A.~\surname{Pilloni}\orcidlink{0000-0003-4257-0928}}
\affiliation{\messina}
\affiliation{\catania}
\author{A.~P.~\surname{Szczepaniak}\orcidlink{0000-0002-4156-5492}}
\affiliation{\jlab}
\affiliation{\ceem}
\affiliation{\indiana}
%

%%%%%%%%%%%%%%%%%%%%%%%%%%%%%%%%%
\begin{abstract}
The construction of general amplitudes satisfying symmetries and $S$-matrix constraints has been the primary tool in studying the spectrum of hadrons for over half a century. In this work, we present a new parameterization, which can fulfill many expectations of $S$-matrix and Regge theory and connects the essential physics of hadron scattering in the resonance region and in asymptotic limits. In this construction, dynamical information is entirely contained in Regge trajectories that generalize resonance poles in the complex energy plane to moving poles in the angular momentum plane. We highlight the salient features of the model, compare with existing literature on dispersive and dual amplitudes, and benchmark the formalism with an initial numerical application to the $\rho$ and $\sigma/f_0(500)$ mesons in $\pi\pi$ scattering.
\end{abstract}
%%%%%%%%%%%%%%%%%%%%%%%%%%%%%%%%%

\maketitle

%%%%%%%%%%%%%%%%%%%%%%%%%%%%%%%%%
\section{Introduction}
\label{sec:introduction}
%%%%%%%%%%%%%%%%%%%%%%%%%%%%%%%%%

In the middle of the twentieth century, a plethora of new strongly interacting particles that decay into pions and nucleons were discovered. To interpret this quickly growing spectrum, Chew, Frautschi, and contemporaries conjectured that these particles must all be bound states of an underlying force analogous to those from non-relativistic potential scattering~\cite{Chew:1961ev,Chew:1962eu}. Crossing symmetry, however, requires that the force carriers of the relativistic theory, i.e., the particles being exchanged, should be those very same bound states and thus these particles must generate themselves through their own interactions. This idea was at the heart of the original bootstrap program, which aimed to examine the structure of relativistic scattering amplitudes in hopes that, when constrained with all fundamental symmetries, a unique self-consistent theory remained~\cite{Chew:1962mpd,Chew:1968pj,Chew:1971xwl}.

Ultimately, the constraints of general $S$-matrix principles~\cite{Eden:1966dnq,Mizera:2023tfe} and discrete symmetries are not enough to define a unique theory of strong interactions~\cite{Castillejo:1955ed,Dyson:1957rgq,Atkinson:1969vec}. It was realized that hadrons are indeed composite particles, but the underlying dynamics are actually of the quarks and gluons within quantum chromodynamics (QCD) and not of a self-generating bootstrap~\cite{Gell-Mann:1962yej,Gell-Mann:1964ewy}. Despite extensive theoretical developments since then~\cite{Maris:2003vk,Scherer:2009bt,Brodsky:2014yha,Shepherd:2016dni,Esposito:2016noz,Guo:2017jvc,Briceno:2017max,Guo:2019twa,JPAC:2021rxu,Mai:2022eur,Gross:2022hyw,Ding:2022ows}, the non-perturbative nature of QCD at low energies means a complete theory of hadronic resonances in terms of those constituents is still an open problem.

More than 50 years since the discovery of QCD, the primary toolkit of hadron phenomenology remains remarkably similar: general amplitudes, satisfying as many symmetries and basic $S$-matrix principles as possible, are constructed to extract meaningful physical information from experimental data~\cite{JPAC:2021rxu}. The key difference is that these amplitudes are now employed to complement and test the predictions from QCD-based approaches such as lattice QCD~\cite{Briceno:2017max,Hansen:2019nir}, functional methods~\cite{Maris:2003vk,Cloet:2013jya,Ding:2022ows}, chiral perturbation theory (ChPT)~\cite{Gasser:1983yg,Gasser:1984gg,Ecker:1994gg,Bernard:1995dp}, and other effective field theories (EFTs)~\cite{Meissner:1987ge,Mai:2022eur,Guo:2017jvc,Neubert:1993mb}.

The specific functional forms used to parameterize amplitudes typically depend on the energies of interest.
In low-energy processes, for example, tools such as dispersion relations have allowed high-precision extraction of pole parameters, while including constraints of unitarity and crossing symmetry at the level of individual partial waves (PWs) (cf.\ Refs.~\cite{Pelaez:2015qba,Hoferichter:2015hva} and references therein). 
At high energies, on the other hand, Regge-based amplitudes, which incorporate an infinite number of PWs, are typically used to describe the phenomenology of peripheral scattering~\cite{Irving:1977ea,Gribov:2009cfk}. 

Because QCD generates both the resonances at low energies as well as the exchanges at high energies, a complete theory of bound states in QCD should still be able to describe both resonance and Regge exchange phenomena, e.g., in the finite energy sum rules~\cite{Dolen:1967jr}. Indeed, Regge inputs are often used to model high-energy contributions to constrain the low-energy PWs (cf., e.g., Ref.~\cite{Pelaez:2015qba}), while resonance information informs the construction of Regge trajectories (RTs), which parameterize high-energy amplitudes (cf., e.g., Ref.~\cite{Huang:2008nr,Sibirtsev:2009bj,Sibirtsev:2009kw}). A ``complete" amplitude, however, which smoothly connects these scattering regimes, has never been satisfactorily established. Such an amplitude would provide a connection between resonances and their properties in the angular momentum plane, which have been argued to give clues to its inner structure~\cite{RuizdeElvira:2010cs,Londergan:2013dza}, and thus give important complementary information to other theoretical approaches.

Historically, attempts at such an all-energies amplitude have fallen into the class of dual models, the most famous of which was proposed by Veneziano~\cite{Veneziano:1968yb}. These amplitudes postulate that the poles in different channels are ``dual" to each other, meaning the sum over poles in one channel will generate the poles in the crossed channel when analytically continued to a different kinematical region~\cite{DelGiudice:1971yjh}. The resulting structure is fundamentally different than the typical construction of crossing symmetric ``interference" or ``isobar" amplitudes, where the poles of each channel are summed coherently such as in Feynman-diagram-based theories~\cite{Jengo:1969fb}. While dual amplitudes offered an appealing and relatively simple connection between low- and high-energy scattering, they were ultimately too restrictive to describe experimental data at any modern level of precision~\cite{Sivers:1971ig}.

State-of-the-art formalisms that do allow high-precision parameterizations, such as those based on Khuri--Treiman~\cite{Khuri:1960zz} equations, work with a truncated set of partial waves. As a result, they no longer incorporate the constraints of analyticity in angular momentum and therefore have uncontrolled behavior at high energies. Usually, this problem is tackled by introducing subtractions, which suppress high-energy contributions and render the necessary dispersion integrals convergent. This requires fixing additional subtraction constants by fitting to data or matching to other theory predictions. These formalisms are thus typically restricted to limited ranges of kinematics and find most applications in meson decays, where the phase space is limited by the mass of the decaying particle. Other formalisms using \mbox{Roy(-like)}~\cite{Roy:1971tc,Garcia-Martin:2011iqs} or Roy--Steiner~\cite{Steiner:1971ms} equations do incorporate an \textit{a priori} known number of subtractions, but require a phenomenological matching between low- and high-energy regions.
Starting from a full amplitude with the correct asymptotic behavior would, in principle, alleviate the need for such subtractions or matching. 

In this article, we revisit the quest for an amplitude that satisfies all necessary constraints and simultaneously describes a wide breadth of scattering phenomena at all energies and scattering angles. We propose a new model for the $2\to 2$ scattering amplitude of spinless particles which:
    \begin{enumerate}
        \item\label{1} is crossing symmetric;
        \item\label{2} is analytic in all energy variables except for cuts at real values;
        \item\label{3} is analytic in angular momentum;
        \item\label{4} satisfies the Froissart--Martin bound~\cite{Froissart:1961ux,Martin:1962rt} and Mandelstam representation with a finite number of subtractions~\cite{Mandelstam:1958xc,Mandelstam:1959bc};
        \item\label{5} has resonances appearing simultaneously as poles in the angular momentum plane and on unphysical sheets of the energy plane;
        \item\label{6} exhibits Regge behavior at high energies with fixed momentum transfer;
        \item\label{7} and exhibits scaling behavior at high energies with fixed scattering angle.
    \end{enumerate}
This is accomplished by the interplay of two pieces: a model for the amplitude itself, which imposes structure expected from $S$-matrix and Regge theory, and a model for the RTs, which feed in the dynamical information of particle properties.  
This two-component formalism allows properties \cref{1,2,3,4,5,6,7} to be satisfied by construction. Unitarity, on the other hand, is not manifestly satisfied and needs to be imposed numerically through the specific implementation of the RTs. As such, we propose a scheme in the same spirit as EFTs, which allows unitarity to be imposed for energies of interest via power counting in momentum barrier factors.

This article is organized as follows: the model for the amplitude constructed from crossing symmetric combinations of ``isobars", i.e., functions that contain the full tower of poles with any spin in a single energy variable. Each isobar is parameterized by RTs appearing in a hypergeometric function as discussed in \cref{sec:hypergeo_iso}. We thereby identify the requirements of the RTs to ensure the model has the desired properties in the limits of interest. In \cref{sec:trajectories}, we discuss the construction of RTs, which satisfy those requirements while being flexible enough to fit data. The utility of our model as a phenomenological tool is demonstrated in \cref{sec:application} by applying it to elastic $\pi\pi$ scattering. We show how our model can be generalized to arbitrary isospin, extract the RTs of the $\rho(770)$ and $\sigma/f_0(500)$ mesons using elastic unitarity, and compare with existing literature. Finally, a summary with discussion of future applications and generalizations is found in \cref{sec:conclusions}.  

To streamline the presentation, several details are relegated to appendices: \Cref{app:hypergeometric_formulae} includes a collection of useful formulae and identities relevant to the properties of the hypergeometric functions. The high-energy limit at fixed scattering angle and the possible connection to inter-meson parton dynamics is explored in \cref{app:large_angle}.  In \cref{app:duality} we discuss general features of dual models and explore how our model fits in with the usual notions of duality.
An explicit demonstration of the crossing properties of the isobar decomposition when generalized to $\pi\pi$ scattering in the isospin limit is provided in \cref{app:isospin}.
Lastly, \cref{app:poles} contains technical details of extracting the pole locations and residues of resonances using this formalism.

%%%%%%%%%%%%%%%%%%%%%%%%%%%%%%%%%%
\section{Hypergeometric isobars}
\label{sec:hypergeo_iso}
%%%%%%%%%%%%%%%%%%%%%%%%%%%%%%%%%

We first consider the elastic scattering of identical, spinless particles 
    \begin{equation}
        \label{eq:s_chan_reaction}
        1\,(p_1) + 2\,(p_2) \to 3 \,(p_3) + 4 \, (p_4) ~,
    \end{equation}
with $p_i^2 = m^2$ and define the usual Mandelstam variables~\cite{Mandelstam:1958xc}
    \begin{align}
        s &= (p_1 + p_2)^2 = (p_3 + p_4)^2 \nonumber  ~,\\
        t &= (p_1 - p_3)^2 = (p_2 - p_4)^2 ~,\\
        u &= (p_1 - p_4)^2 = (p_2 - p_3)^2 \nonumber ~,
    \end{align}
which satisfy the on-shell condition $s + t + u = 4m^2$.  Because the particles are identical, crossing symmetry requires the single function $\Amp(s,t,u)$, denoting the scattering amplitude of the reaction \cref{eq:s_chan_reaction}, to also simultaneously describe the reactions $1 + \bar{3} \to \bar{2} + 4$ and $1 +\bar{4} \to 3 + \bar{2}$ through the analytic continuation of momenta, or equivalently the interchange of Mandelstam variables.  

The amplitude is expressible as an infinite sum of PWs in a given channel. For instance, we may write~\cite{Martin:1970hmp,Collins:1977jy} 
    \begin{equation}
        \label{eq:general_PW_expansion}
        \Amp(s,t,u) = \sum_{j=0}^\infty (2j+1) \, P_j(z_s) \, \PW_j(s) ~,
    \end{equation}
in terms of the cosine of the $s$-channel scattering angle $z_s$ and the $s$-channel PWs~\cite{Jacob:1959at},
    \begin{equation}
        \label{eq:general_PW}
        \PW_j(s) = \frac{1}{2}\int^1_{-1} \dd z_s \,P_j(z_s) \, \Amp(s, t(s,z_s), u(s,z_s)) ~,
    \end{equation}
where $P_j(z_s)$ are the Legendre polynomials.
The crossed channel variables are related to the angular variable $z_s$ via
\begin{subequations}
    \label{eq:t_and_u_from_zs} 
    \begin{align}
        t(s,z_s) &= - 2 \, q^2_s \,(1 - z_s)~, \\
        u(s,z_s) &= - 2 \, q^2_s \,(1 + z_s)~,       
    \end{align}
\end{subequations}
where $q_s = \sqrt{s-4 \, m^2}/2$ is the modulus of the $3$-momentum of the initial or final state particles in the $s$-channel center-of-mass (CM) frame. \Cref{eq:t_and_u_from_zs} can be inverted to express the angular variable in terms of the Mandelstam variables:
    \begin{equation}
        \label{eq:zs_from_tu}
        z_s = \frac{t -u}{4 \, q_s^2} ~.
    \end{equation}

The PWs in \cref{eq:general_PW} are analytic functions of energy, which contain branch cuts from the lowest multi-particle threshold, $s = 4 \, m^2$ to infinity on the right-hand side of the complex $s$-plane and another from $s = 0$ to negative infinity on the left, referred to as right-hand cut (RHC) and left-hand cut (LHC), respectively. The RHC singularities arise from unitarity in the direct channel, in this case the $s$-channel, while the LHC is related to the dynamics of the crossed channels. Because the dependence on $t$ and $u$ enters only through the angular polynomials, the two-cut structure and the infinite number of terms in \cref{eq:general_PW_expansion} are required to reconstruct the full amplitude and satisfy crossing symmetry. Specifically, while \cref{eq:general_PW} is a general definition of the PW amplitude for any $s$, the expansion in \cref{eq:general_PW_expansion} will only converge within a limited region of the complex $z_s$-plane (equivalently the $t$- or $u$-planes) given by the Lehmann ellipse~\cite{Lehmann:1958ita,Martin:1965jj}. Recovering the crossed channel PW expansion, i.e, \cref{eq:general_PW_expansion} in terms of $t_j(t)$, thus involves a re-summation of the infinite sum outside this radius of convergence.

To simplify the introduction of the model, we restrict ourselves to the case of isoscalar particles and will generalize to isovectors in \cref{sec:application}.
In an isobar or interference model, the full amplitude is assumed to decompose into a crossing symmetric sum of terms describing individual scattering channels~\cite{Jengo:1969fb}:
    \begin{equation}
        \label{eq:gen_isobar_decomp}
        \Amp(s,t,u) = \mF{+}(s, z_s) + \mF{+}(t, z_t) + \mF{+}(u,z_u) ~.
    \end{equation}
The superscript refers to the required symmetry of each term with the interchange of final state particles, i.e., we require $\mF{\pm}(s,z_s) = \pm \, \mF{\pm}(s,-z_s)$. In the identical particle case all three channels enter with the same function $\mF{+}(x, z_x)$ for $x = s$, $t$ or $u$, and it is straightforward to verify that \cref{eq:gen_isobar_decomp} is fully invariant under the interchange of any two Mandelstam variables, thereby satisfying crossing symmetry.

$\mF{\pm}(s,z_s)$ is the $s$-channel isobar, which is assumed to contain poles only in $s$. Similarly, the isobars $\mF{\pm}(t,z_t)$ and $\mF{\pm}(u,z_u)$ only contain poles in $t$ and $u$, respectively, and enter with the cosines of the crossed channel scattering angles $z_t$ and $z_u$. These angles are given by \cref{eq:zs_from_tu} with the interchange of $(s\leftrightarrow t)$ and $(s\leftrightarrow u)$, respectively.

In the $s$-channel physical region, then, the direct channel isobar, i.e., $\mF{\pm}(s,z_s)$, produces the pole structure associated with the resonance spectrum. The other two terms produce a smooth background associated with the exchanges of those same resonances in the crossed channels. This separation of poles in each variable means that the amplitude is primarily driven by the first term in \cref{eq:gen_isobar_decomp} at low energies where $s$-channel resonances are observed. At high energies, it is instead dominated by the other two terms where the characteristic Regge behavior must emerge from the crossed channel.

For the construction in \cref{eq:gen_isobar_decomp} to describe both poles and exchange behavior in the appropriate limits simultaneously, we parameterize the dynamics of each channel in terms of RTs. These are dynamical functions, which encode the properties of not only single resonances, but potentially infinite towers of particles at all energies by interpolating their pole locations in the complex angular momentum plane~\cite{Regge:1959mz}. Guided by this, we decompose the isobars as
    \begin{equation}
        \label{eq:mF_pm}
        \mF{\pm}(s,z_s) = \sum_i \frac{g^2_i}{2}\left[F(\alpha_i(s), \,\nu_s) \pm  (\nu_s \leftrightarrow -\nu_s)  \right] ~,
    \end{equation}
where $\alpha_i(s)$ is a RT and $g_i$ is a real coupling constant. The sum runs over all RTs, which are expected to contribute to the process. 
Each RT in the sum corresponds to a family of hadrons with increasing spin and the same quantum numbers. The natural parity, $C$-odd isovectors, for instance, i.e., the $\rho(770)$, $\rho_3(1690)$, etc., are all expected to lie on one trajectory, while isoscalars such as the $\sigma/f_0(500)$ and its possible excitations would be described by another RT. The properties and explicit construction of RTs will be explored in detail in \cref{sec:trajectories}, but for now it is sufficient to assume that each is an analytic function of $s$ except for a RHC required by unitarity~\cite{Collins:1977jy}. 

The angular dependence enters through the crossing variable $\nu_s$, which depends only on crossed channel Mandelstam variables
    \begin{equation}
        \nu_s = \qzs = \frac{t-u}{4}~.
    \end{equation}
The two terms with respect to $\pm \nu_s$ manifestly satisfy the required symmetry with respect to $(z_s \leftrightarrow -z_s)$.

Finally, we define each term in \cref{eq:mF_pm} corresponding to the exchange of a single RT with the form
    \begin{widetext}
    \begin{equation}
        \label{eq:f(alpha_z)}
        \fas =  \Gamma(\jmin - \as) \, \hat{\nu}_s^\jmin \, \gaussFR{\jmin+1, \jmin - \as}{\jmin + 1 - \as}{\hat{\nu}_s}
        ~,
    \end{equation}
    \end{widetext}
in terms of the $\Gamma$ function and the regularized hypergeometric function $\gFR$ defined from the usual one by
    \begin{align}
        \label{eq:gaussFR}
        \gaussFR{a, b}{c}{d}&=\frac{\gaussF{a,b}{c}{d}}{\Gamma(c)} ~.
    \end{align}
The regularized hypergeometric function\footnote{General formulae regarding hypergeometric functions, which are relevant to our results, are collected in \cref{app:hypergeometric_formulae}.} is complex analytic with no pole singularities at finite values of its arguments and with a branch cut starting when its last argument is $d=1$ and extending to infinity so long as $a$ or $b$ are not negative integers. We introduce a momentum scale $\Lambda$ and define $\qhats^2 \equiv q_s^2 / \Lambda^2$ and $\hat{\nu}_s=\nu_s/\Lambda^2$. In \cref{sec:regge}, this scale will be identified as the characteristic scale of Regge physics. 
The remaining parameter $\jmin \geq 0$ is an integer, which corresponds to the lowest spin of a physical particle lying on the RT, e.g., the $\sigma/f_0(500)$ that has $\jmin=0$ or the $\rho$ whose trajectory has $\jmin=1$. 

The usefulness of the hypergeometric functions in modeling scattering amplitudes is well known, since the hypergeometric function provides an analytic continuation of angular polynomials beyond integer spins and therefore hints at a natural connection with Regge physics. Because of this, isobar models exploiting the analytic properties of hypergeometric functions were proposed as early as the seventies as an alternative to dual models~\cite{Donnelly:1972dnn,Botke:1972xn}. Hypergeometric functions also appear naturally in the modeling of multi-Regge processes~\cite{Brower:1974yv,Shimada:1978sx}. More recently, they were shown to be integral in elucidating the role of Reggeized pion exchange in high-energy $\pi^\pm$ photoproduction via analytic continuation in angular momentum~\cite{Montana:2024ngi}.
Thus, before looking at specific features of \cref{eq:f(alpha_z)}, we first consider its general analytic structure brought upon by the hypergeometric function.

Since both $\gFR$ and $\Gamma$ are analytic functions in all their arguments, $\fas$ is analytic in both the RT and crossed channel variables. Further, at fixed $t$, because the $s$-dependence only implicitly through the RT (which is assumed to be analytic everywhere except for the RHC), the isobar is simultaneously analytic in both $s$ and $\as$. At fixed angle, additional dependence on $s$ enters through $q_s^2$, which is polynomial, and therefore introduces no further singularities. As the other isobars, which enter in the crossed channel terms, are related to $\fas$ by the interchange of Mandelstam variables, the full amplitude \cref{eq:gen_isobar_decomp} will be analytic in all three energy variables, and the RT appearing in all three channels. Since the RT will be shown to be intimately related to spin, this property will be integral to the analyticity of the amplitude both in energy and in complex angular momentum.

If $\as$ has a RHC,\footnote{In principle, $\as$ should contain many branch points corresponding to every multi-particle threshold that the RT can couple to.} then $\fas$ will also have a RHC. The $\Gamma(\jmin - \as)$ in \cref{eq:f(alpha_z)} introduces poles at integer values of $\as$ and will be discussed at length in \cref{sec:resonances}. Similarly, the crossed channel isobars with respect to $\fat$ contain poles in $t$ and will develop LHCs associated with the RHC of $\at$ as viewed from the $s$-channel physical region. 

\begin{figure}[t]
        \centering
        \includegraphics[width=0.48\textwidth]{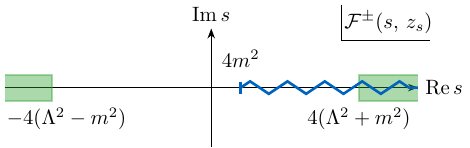}
        \includegraphics[width=0.48\textwidth]{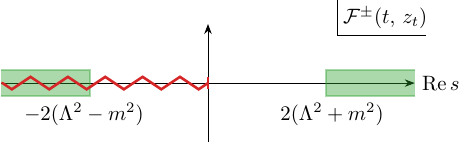}
        \caption{The cut structure of $\mF{\pm}(s,z_s)$ (top) and $\mF{\pm}(t,z_t)$ (bottom) in the complex $s$-plane for fixed $z_s=1$. The zigzags denote cuts induced by the RTs. The RHC (blue) starting at threshold originates from $\as$ and the LHC (red) from $\at$ in the crossed channel. The green shaded regions denote where the secondary cuts of the regularized hypergeometric function begin to overlap those of the RTs.}
        \label{fig:cut_structure_isobars}
    \end{figure}

The aforementioned cut structure of the hypergeometric function will then split the behavior of the isobar into two distinct regimes below and above the secondary branch cut openings at each $\hat{\nu}_{s,t,u} = 1$.
Specifically, because $\hat{\nu}_s$ involves the product of momentum and the scattering angle, the hypergeometric function contributing to the direct channel, \cref{eq:f(alpha_z)}, has a branch point, which depends on the value of $z_s$. Although this cut is unphysical, it will only affect how unitarity can be imposed. Provided $\Lambda$ is large enough, the effects of these cuts are irrelevant in the resonance region.  The values $z_s = \pm 1$ generate the lowest lying branch points, which occur at $q^2_s = \pm\Lambda^2$ or equivalently at $s = \pm 4(\Lambda^2 \pm m^2)$. The symmetrized isobar $\mF{\pm}(s,z_s)$ will feature both of these cuts as shown diagrammatically in \cref{fig:cut_structure_isobars}. 

Because crossing symmetry requires that the same scale $\Lambda$ also enters the crossed channel isobars, we must calculate the location of the branch points of $\mF{\pm}(t,z_t)$ with respect to $\hat{\nu}_{t} = 1$ as a function of $s$ and $z_s$. These can be found to appear at $s = \pm 2(\Lambda^2 \pm m^2)$ and therefore are lower lying than the secondary cuts of the direct channel. The cuts of the $u$-channel terms are identical. The cut structure of the crossed channel isobars in the $s$-plane is also illustrated in \cref{fig:cut_structure_isobars}. This two-cut structure of each isobar above a characteristic scale mimics that of the full amplitude, and will ensure that analyticity and unitarity constraints are still satisfied when Reggeized. 

\begin{figure}[t]
        \centering
        \includegraphics[width=0.48\textwidth]{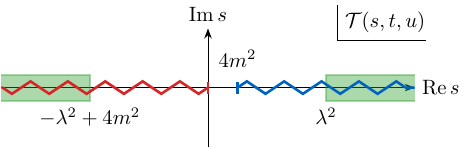}
        \caption{The cut structure of the full amplitude $\Amp(s,t,u)$ in the complex $s$-plane for fixed $z_s=1$. Color scheme for the cuts as in \cref{fig:cut_structure_isobars}.}
        \label{fig:cut_structure_fullamplitude}
    \end{figure}

The first secondary cut in the $s$-channel physical region will thus be the right-hand side cut coming from the hypergeometric function in the crossed channel isobars and we define 
    \begin{equation}
        \lambda^2 = 2(\Lambda^2+m^2)~.
    \end{equation}
The energy $s = \lambda^2$ is thus the maximal energy at which all terms in \cref{eq:gen_isobar_decomp} lie below the branch points of the hypergeometric function for the entire physical range of the scattering angle in all three Mandelstam variables. Below this energy, each term only contains the RHC from the RT in its respective energy variable and the model resembles the structure of an isobar model in the Khuri--Treiman (KT) dispersion representation~\cite{Khuri:1960zz,Bronzan:1963mby,Aitchison:2015jxa,Albaladejo:2019huw}. A diagrammatic representation of the cut structure for the full amplitude (i.e., the crossing symmetric sum of individual isobars) in the $s$-plane is shown in \cref{fig:cut_structure_fullamplitude}. The same cut structure will also appear in the $t$- and $u$-channel physical regions, respectively, due to crossing symmetry.

For energies $s > \lambda^2$, at least one isobar is evaluated above the additional branch points and the structure becomes more complicated. As we will explore in the following subsections, this transition point marks the energy at which the PW expansion diverges and must be re-summed into something that is Regge-behaved. In this way, these two kinematic regions reproduce the near-threshold resonance and asymptotic Regge regimes, respectively, which are connected via the complete analyticity of our isobars. 

%%%%%%%%%%%%%%%%%%%%%%%%%%%%%%%%%%
\subsection{Regge region \texorpdfstring{$(s\gg \lambda^2)$}{}}
\label{sec:regge}
%%%%%%%%%%%%%%%%%%%%%%%%%%%%%%%%%
We first consider the behavior of the isobars for energies above $\lambda^2$. For simplicity, we take $s \gg \lambda^2$ to study the Regge limit, i.e., $s,-u\to\infty$ with $t < 0$ fixed. A general crossing symmetric amplitude with the correct analytic properties is known to manifest Regge behavior in this limit and decomposes into a sum of terms of the form~\cite{Regge:1959mz,Collins:1977jy,Gribov:2009cfk} 
    \begin{equation}   
        \label{eq:regge_behavior} 
      \mathbb{R}(s,t) = \beta(t) \, \xi_\pm(t) \, \Gamma(\jmin - \alpha(t)) \left(\frac{s}{s_0}\right)^{\alpha(t)} ~.
    \end{equation}
This is the quintessential form of a Reggeon exchange, where $\beta(t)$ is an arbitrary real function of $t$ referred to as the Regge residue. The signature factor, $\xi_\pm(t)$, is an oscillatory function of the form
    \begin{equation}
        \label{eq:signature_factor}
         \xi_\pm(t) = \frac{1}{2}\left[\pm1 + e^{-i\pi\at}\right] ~,
    \end{equation}
with the $\pm1$ denotes the signature of the RT $\alpha(t)$.
The characteristic power-law behavior $s^\at$ is associated with moving pole singularities, i.e., as a function of $t$, in the complex angular momentum plane and arises from the re-summation of leading powers of $s$ in the angular polynomials of the PW expansion, \cref{eq:general_PW_expansion}, outside its radius of convergence.

The Froissart--Martin bound~\cite{Froissart:1961ux,Martin:1962rt} limits the possible indefinite growth of the amplitude and restricts $\alpha(0) \leq 1$ for all RTs. The bound is saturated by the Pomeron with $\alpha_\mathbb{P}(0) \simeq 1$, while all other RTs corresponding to hadrons must be subleading, e.g., the $\rho$ is found to have $\alpha_{\rho}(0) \simeq 0.5$~\cite{Barnes:1976ek}.
Crossed channel unitarity fixes the imaginary part of the amplitude in this limit to be given by the signature factor.

We will show that the full amplitude constructed in \cref{eq:gen_isobar_decomp} manifests the asymptotic behavior of \cref{eq:regge_behavior}, and thus the $\alpha(t)$ that appears in \cref{eq:f(alpha_z)} is indeed a RT in the usual sense. Further identification of the RT as poles in the $j$-plane will be explored in the next subsection by demonstrating that the same function also generates resonances, i.e., poles in the PW amplitude. 

We begin by considering the $t$-channel isobar, $\mF{\pm}(t, z_t)$. At fixed $t <0$, i.e., below the RHC, $\at$ is real and finite. Thus, taking the Regge limit only entails considering the hypergeometric function with $|\hat{\nu}_t| \to \infty$. 
Using \cref{eq:gaussFR_times_d_limit} we must consider two cases for $\at >-1$ and $\at\leq -1$ when taking this limit, cf.\ \cref{app:hypergeometric_formulae}. If $\at > -1$ the isobar can be written as
    \begin{align}
        \label{eq:t_limit_gm1}
        F(\alpha(t) > -1, \nu_t \to \infty) & \\
        &\hspace{-2.4cm} =\left[\frac{\Gamma(1+\at)}{(-1)^\jmin \jmin!}\right] \, \, \Gamma(\jmin-\at)   \left(\frac{u-s}{4\Lambda^2}\right)^{\at} ~, \nonumber
    \end{align}
which resembles \cref{eq:regge_behavior} without the signature factor. We already notice the emergence of a $\Gamma(\jmin - \at)$ factor, which would generate poles at \textit{positive} integers $\at = j \geq \jmin$. Because we are explicitly considering $t<0$, which is below the RHC, these poles are never manifested and requiring $\jmin > \alpha(0)$ forbids the possibility of negative energy poles appearing in the physical region.

We additionally note that the $\Gamma(1+\at)$ factor appears to generate poles at \textit{negative} integers $\at = j \leq -1$. These poles, however, do not exist as, if $\at < -1$, the amplitude is instead given by the second term in the asymptotic expansion in \cref{eq:gaussFR_times_d_limit}:
    \begin{align}
        \label{eq:t_limit_lm1}
        F(\at < -1, \nu_t \to \infty) =
        \frac{(-1)^\jmin}{-1-\at} \, \left(\frac{u-s}{4\Lambda^2}\right)^{-1} ~. 
    \end{align}
At the transition point $\at = -1$, the hypergeometric function can be computed explicitly and shown to be finite with energy behavior of $s^{-1} \, \log s$ smoothly connecting the two limits in \cref{eq:t_limit_gm1,eq:t_limit_lm1}, cf.\ \cref{eq:gaussFR_times_atm1}. 
 
The behavior of \cref{eq:t_limit_lm1} means that at large $s$, the amplitude is always bounded from below by a fixed power and therefore satisfies the Cerulus--Martin bound~\cite{Cerulus:1964cjb}. The specific form of \cref{eq:t_limit_lm1} is actually expected from Regge theory as unitarity prohibits amplitudes that only contain poles in the $j$-plane from falling faster than $s^{-1}$~\cite{Gribov:1962gb,Gribov:1965hg} and can be attributed to the condensation of all Regge poles at $j = -1$ in the left-half angular momentum plane~\cite{Gribov:2003nw,Gribov:2009cfk}. The asymptotic behavior of scattering amplitudes at fixed angles, that is, additionally taking $t \to -\infty$ with the ratio $s/t$ kept finite, is closely related to the behavior in \cref{eq:t_limit_lm1} and has been proposed to be connected with the microscopic dynamics of the partons exchanges. This connection is explored further in \cref{app:large_angle}.

Inserting \cref{eq:t_limit_gm1} into \cref{eq:mF_pm}, we see that Bose symmetry with respect to $z_t$ generates the signature factor and, using $s\sim - u$, yields a leading Regge behavior of
    \begin{align}
        \label{eq:F_tlimit}
        &\frac{1}{2}\left[\fat \pm \fatm\right] 
         \\
        &\quad =\left[\frac{\Gamma(1+\alpha(t))}{(-1)^\jmin \jmin!}\right] \, \xi_\pm(t)  \,\Gamma(\jmin - \alpha(t)) \, \left(\frac{s}{2\Lambda^2}\right)^{\alpha(t)}   ~,    
        \nonumber
    \end{align}
so long as $\at$ is not too negative.

Reading off the Regge residue by comparing \cref{eq:gen_isobar_decomp,eq:mF_pm,eq:F_tlimit} with \cref{eq:regge_behavior} gives \mbox{$\beta(t) = g^2 \, \Gamma(1+\alpha(t))/((-1)^\jmin \jmin!)$} and the characteristic Regge scale $s_0 = 2 \, \Lambda^2$. Note that this scale corresponds to the location of the secondary branch point $s_0 \approx \lambda^2$ in the limit $\Lambda^2 \gg m^2$. Because $g$ and $\at$ are both real, the imaginary part of \cref{eq:F_tlimit} emerges solely from the signature factor $\xi_\pm(t)$ as expected from $t$-channel unitarity. 

As we have shown, Regge behavior emerges in the $t$-channel isobar when taking $t <0$ finite. Due to crossing symmetry, the same Regge behavior is found in all other similar limits, e.g., the $s$-channel isobar Reggeizes when $s <0$  with $t,-u \to \infty$. However, precisely because all isobars contribute to the full amplitude, it is necessary to also show that all other isobars vanish faster than \cref{eq:F_tlimit}. 

Because in the Regge limit $s, - u\to+\infty$, we must consider the behavior of the RTs at infinity. We will assume that the RTs are unbounded in both directions as this will ensure that the behavior of \cref{eq:t_limit_gm1,eq:t_limit_lm1} is always the leading power of $s$ at finite $t$.

Looking at the $u$-channel term, because $u \to -\infty$ is below the RHC, $\au$ is real and we require that $\alpha(u\to-\infty) \to -\infty$.
If this is the case, the trajectory will eventually cross $\au = -1$ and the isobar will behave as 
    \begin{equation}
        \label{eq:u_limit}
        F(\au \to -\infty, \nu_u \to -\infty) = \frac{(-1)^\jmin}{\au}\left(\frac{u}{4\Lambda^2}\right)^{-1} ~. 
    \end{equation}
The factor of $\au^{-1} \to 0$ ensures that this term will vanish faster than the leading Regge term at any $t<0$.\footnote{Satisfying this limit forbids considering the coupling $g$ in \cref{eq:mF_pm} as a function of energy. If $g$ is entire and $g<|\au|<u$ then $g$ is constant by Liouville's theorem.} Computing the same limit for the other term appearing in \cref{eq:mF_pm}, i.e., with $(\nu_u \leftrightarrow -\nu_u)$,  proceeds identically.

The last term to consider is the $s$-channel isobar as $s \to \infty$ and thus involves considering complex $\as$ above the RHC. 
To demonstrate Regge behavior it is sufficient to assume that the RT is unbounded, as before, but grows slower than $s$. While stronger bounds are possible, less-than-linear growth can be shown to be a requirement of RTs based on the most general analyticity principles~\cite{Gribov:1962gb} and is sufficient at this stage. With this in mind, we will identify additional requirements from our model on the asymptotic behavior of $\as$ to be considered in \cref{sec:trajectories}. 
Because of the different power-law behavior in \cref{eq:t_limit_gm1,eq:t_limit_lm1}, we need to consider the two different cases of $\Re\as \to \pm \infty$ asymptotically.

If $\Re\as \to -\infty$, the limit will be identical to that of the $u$-channel in \cref{eq:u_limit} and will vanish faster than Regge behavior regardless of $\Im\alpha(s)$.

The more nuanced limit arises from RTs that rise indefinitely to positive infinity. Evaluating the Regge limit in this case yields an asymptotic behavior of the form
    \begin{align}
        \label{eq:s_limit_plus}
         F(\as\to\infty, \pm\nu_s \to \pm \infty)& \notag\\
        &\hspace{-2cm} =\frac{\alpha(s)^\jmin}{\jmin!} \,\frac{-\pi}{\sin \pi \as} \, \left(\frac{\mp s}{4\Lambda^2}\right)^{\as}~,
    \end{align}
where we have used the Euler reflection formula and kept only the leading powers of the Pochhammer symbol, \cref{eq:pochhammer}, at large arguments, i.e., $(x)_j \to x^j$. 

Since $\as$ is complex, we may separate the contributions from its real and imaginary parts such that, up to overall constants, the modulus of \cref{eq:s_limit_plus} is given by
    \begin{align}
        \label{eq:s_limit}
        &\left|\frac{\alpha(s)^\jmin}{\sin\pi\as} \left(\frac{s}{4\Lambda^2}\right)^{\as}\right|  \\
        &\hspace{1.5cm} \propto\exp \left[\Re\as \log (s/4\Lambda^2) - \pi \Im\as\right] ~.
        \nonumber
    \end{align}
The case involving $(-s)^\as$ is subleading and not shown explicitly. 
Because we assume $|\as| < s$ asymptotically, terms proportional to $\log|\as|$ in the exponential are neglected. In this form, the $s$-channel isobar is exponentially suppressed if we require the RT to asymptotically satisfy
    \begin{equation}
        \label{eq:Imalpha_bound}
        \Im \as > \frac{1}{\pi} \Re\as \log\left( \frac{s}{4\Lambda^2}\right) ~.
    \end{equation}
Such a condition is not new and naturally emerges in Regge-behaved models with complex, rising RTs to ensure the amplitude is polynomial-bounded at infinity~\cite{Botke:1972xn,Childers:1968vnm,Bugrij:1973ph,Trushevsky:1975yf,Degasperis:1970us,Botke:1969xv}. In particular, Ref.~\cite{Botke:1972xn} shows that a model and RT satisfying \cref{eq:s_limit_plus,eq:Imalpha_bound}, respectively, are sufficient for the Mandelstam double dispersion representation to converge without subtractions. This property will guarantee that the amplitude has no essential singularities at infinity in any direction in the (complex) Mandelstam plane.

 We note that \cref{eq:u_limit} is trivially satisfied by \cref{eq:Imalpha_bound} and thus is enough for both $\mF{\pm}(s,z_s) \to 0$ and $\mF{\pm}(u,z_u) \to 0$ faster than the Regge behavior of $\mF{\pm}(t,z_t)$ at arbitrary $t<0$. Thus with a reasonably well-behaved $\as$ whose imaginary part grows sufficiently fast, the full crossing symmetric amplitude \cref{eq:gen_isobar_decomp} will be properly Regge-behaved in all channels.

%%%%%%%%%%%%%%%%%%%%%%%%%%%%%%%%%
\subsection{Resonance region \texorpdfstring{($s< \lambda^2$)}{}}
\label{sec:resonances}
%%%%%%%%%%%%%%%%%%%%%%%%%%%%%%%%%
The Regge behavior explored in the previous subsection is related to the presence of particles exchanged in the crossed channels at high energies. Crossing symmetry dictates that these same particles must be present in all other channels and, therefore, manifest as resonances when the CM energy is near the pole location. We explore the behavior of our isobar, \cref{eq:f(alpha_z)}, in the region below the secondary branch cuts, which manifests these resonant poles.

By construction, for $4m^2 \leq s< \lambda^2$, we have \mbox{$0\leq |\hat{\nu}_s|< 1$} and the isobar \cref{eq:f(alpha_z)} can be expanded in powers of $\hat{\nu}_s$ with \cref{eq:gaussFR_power}. For the discussion of poles in this section, we switch to using $\hat{\nu}_s = \qhatzs$ to more easily identify the angular structure and write
    \begin{align}
        \label{eq:fas_poles}
        F(\as, \, \nu_s) &=\sum_{j=\jmin}^\infty \, n_j \, \frac{(\qhatzs)^{j}}{j-\as}~,
    \end{align}
with $n_j \equiv j! / \left[\jmin! \, (j-\jmin)!\right]$.
Clearly, \cref{eq:fas_poles} is the sum of simple poles whose residues are polynomials in the crossed channel variables. The analytic structure of \cref{eq:fas_poles} is thus entirely determined by that of $\as$. In the absence of bound states, having no poles on the physical Riemann sheet means any point $s_j$ satisfying \mbox{$(j - \alpha(s_j)) = 0$} must appear underneath the RHC of the trajectory in the lower-half complex $s$-plane.

To interpret the residue of these poles, we decompose the monomial in $z_s$ into Legendre polynomials as
    \begin{align}
        \label{eq:monomial}
        z_s^j
        &
        = \sum_{\jp=0}^j (2\jp+1) P_\jp(z_s) \, \mu_{j\jp}\, \tau_{j+\jp}~,
    \end{align}
with $\mu_{j\jp} \equiv  j!/\left[(j-\jp)!! \, (j+\jp+1)!!\right]$. Since the monomial has definite parity with respect to $z_s$, the only non-zero terms in the expansion involve Legendre polynomials of the same parity as $j$ and we can define the parity symbol \mbox{$\tau_{k} \equiv \left[1+(-1)^k\right]/2$}, which is $0$ and $1$ for odd and even indices, respectively. 

\begin{figure}[t]
        \centering
        \includegraphics[width=0.48\textwidth]{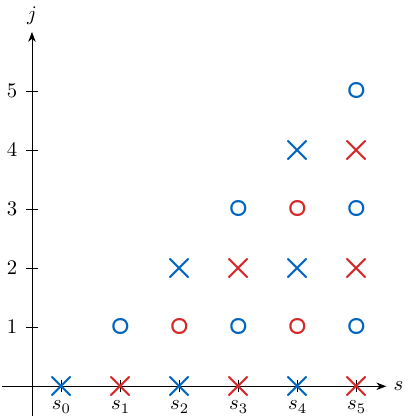}
        \caption{Chew--Frautschi plot illustrating the schematic pole structure of \cref{eq:mF_pm} with a generic, indefinitely rising trajectory. Isobars will have poles at each energy $s_j$ in the complex plane satisfying $\alpha(s_j) = j$.  The horizontal axis thus schematically represents \textit{complex pole locations} that do not require the RT to be real or linear.
        We show poles from two terms both with $\jmin=0$: the leading trajectory $\as$ (blue) and the first daughter trajectory $\as - 1$ (red). Crosses (circles) are poles appearing in the isobars with even (odd) signature.}
        \label{fig:chew_frautschi}
    \end{figure}

Combining \cref{eq:fas_poles,eq:monomial}, as $\as\to j$, a single term will dominate and can be written as
    \begin{align}
        \label{eq:fas_single_pole}
        F(\alpha(s)\to j, \, \nu_s)  
        \\ 
        % &\hspace{-1.5cm}=\sum_{\jp = 0}^j (2\jp+1) \, \qhats^{2j} \,  P_\jp(z_s) \, \left[ \frac{\tau_{j+\jp} \, n_j \, \mu_{j\jp}}{j-\as} \right] ~.
        % \notag  
        % \\
        &\hspace{-1.5cm}= \frac{\qhats^{2j} }{j-\as} \left[\sum_{\jp = 0}^j (2\jp+1) \, \,  P_\jp(z_s) \,  n_j \, \tau_{j+\jp} \, \mu_{j\jp} \right] ~.
        \notag 
    \end{align}
Here we see the pole at $\alpha(s_j) = j$ describes not only a single resonance of spin-$j$, i.e., the residue is proportional to $P_j(z_s)$, but also resonances of all possible same-parity spins $\jp < j$ at the same mass. This coincidence of masses for particles of decreasing spin is observed in the \mbox{Chew--Frautschi} plots of many meson families~\cite{Chew:1962eu} and referred to as the spectrum of ``daughter poles"~\cite{Collins:1977jy,Anisovich:2000kxa}. These are typically understood as resonances lying on RTs parallel to $\as$ but shifted down by integer units of angular momentum, e.g., the $k$-th daughter appears on the trajectory \mbox{$\as - k$}. The resulting pole structure of the isobars is illustrated by the \mbox{Chew--Frautschi} plot in \cref{fig:chew_frautschi}. It is important to note that the spectrum resembles that of the Veneziano model~\cite{Sivers:1971ig,Szczepaniak:2014bsa}, however, as mentioned above, the pole must be located in the complex $s$-plane.

If a pole is sufficiently narrow (i.e., with \mbox{$\Re s_j \gg \Im s_j$}) and isolated, the trajectory may be expanded around the projection of the pole position on the real axis recovering a \mbox{Breit--Wigner (BW)} form~\cite{Breit:1936zzb,Gribov:2003nw,JPAC:2018zjz}. In principle, however, arbitrary lineshapes can be implemented with an appropriately chosen parameterization for $\alpha(s)$. 

The structure of \cref{eq:fas_poles} is robust for complex and non-linear $\as$, with the residue of each term always appearing as a fixed-order polynomial of the crossed channel variables and thus avoids the appearance of ancestor poles~\cite{Paciello:1969ry}. Furthermore, the ``parent pole", i.e., the term with $\jp = j$, always appears with the correct angular momentum barrier factor $q_s^{2j}$ as required to remove kinematic singularities in the angular variable and the factorization of Regge pole residues~\cite{Gribov:1965hg,Gribov:2009cfk}. The same barrier factor multiplies daughter poles, which will also be free of kinematic singularities, but vanish faster at threshold than required by analyticity. Since any number of trajectories can be added in \cref{eq:mF_pm}, this problem can be remedied by adding the daughter trajectory itself into the sum, i.e., including terms with \mbox{$F(\alpha(s) -k, \nu_s)$}.

Because every pole term in \cref{eq:f(alpha_z)} has a definite parity, when inserted into \cref{eq:mF_pm}, Bose symmetry acts as a filter with the even-signature combination 
    \begin{equation}
        \label{eq:mF_resonance}
        \mF{+}(s,z_s)  =\sum_i \,  \sum_{j=\jmin}^\infty \,\tau_j \, \left[    g^2_i \, n_{ij}  \,  \frac{ (\qhatzs)^j}{j-\alpha_i(s)}\right]
    \end{equation}
only containing poles at even integers.\footnote{Because $n_{ij}$ depends on $\jmin$ it implicitly depends on $\alpha_i(s)$. This is the same quantity as in  \cref{eq:fas_poles} where, since only one RT was considered, the index $i$ was omitted.} 
Similarly, the anti-symmetric combination $\mF{-}(s,z_s)$ will only have poles at odd values of $j$, i.e., with a factor of $\tau_{j+1}$. In this way, $\tau$ is the restriction of the signature factor in \cref{eq:signature_factor} to integer values of angular momentum, and enforces that only particles of a definite parity/signature appear on each RT. 

Turning to the $t$-channel isobars, since $|\hat{\nu}_t| < 1$, we find the identical structure to \cref{eq:mF_resonance}, but with \mbox{$(s\leftrightarrow t)$}. Therefore, the dependence of $s$ will be a polynomial and the only possible cut comes from $\at$. Since in the $s$-channel physical region, $t\leq0$ is below the RHC of $\at$, the entire contribution from \mbox{$\mF{\pm}(t,z_t)$} is a real and smooth background to the direct channel. The $u$-channel isobar follows analogously. 

Combining all three isobars to the full amplitude we can conclude that in the $s$-channel physical region, with \mbox{$4m^2\leq s< \lambda^2$}, \cref{eq:gen_isobar_decomp} will only have poles in $s$ and the imaginary part arises only by the $s$-channel isobar
    \begin{align}
        \label{eq:ImA_resoances}
        \Im \Amp&(s,t,u) = \Im \mF{+}(s, \, z_s) \\
        &= \sum_i  \left[\sum_{j=\jmin}^\infty \frac{ n_{ij} \,\tau_j\, (\qhatzs)^j}{|j-\alpha_i(s)|^2} \right] \, g^2_i \,\Im \alpha_i(s) ~. \nonumber
    \end{align}
Although we have ignored unitarity thus far, we see that the imaginary part of the amplitude (and therefore of the PWs) comes from the interplay of the different RTs. As the RTs are, in principle, complicated functions correlating information of all PWs and inelastic channels simultaneously, it is likely impossible to construct a finite set of trajectories that exactly unitarize \cref{eq:ImA_resoances} at all energies.
However, since the term in the brackets is a convergent expansion in momentum (divided by the energy scale $\Lambda$),\footnote{Since there is additional $s$-dependence in the RT, this is not a true expansion in momentum. Instead, this is analogous to the suppression of higher PWs below the radius of interaction~\cite{Gribov:2009cfk}.} one can impose unitarity up to some energy of interest by power counting factors of $\qhats^2$ as we will show in \cref{sec:application}. 

A similar structure to \cref{eq:ImA_resoances} emerges in the KT formalism, i.e., the discontinuity of the full amplitude along the RHC is also that of the direct channel isobar. 
One crucial difference with a typical KT decomposition is that the sum in \cref{eq:fas_poles} is necessarily infinite and each isobar will contribute to all allowed PWs owing to its analyticity in $j$. In this energy range, the series in $\qhats$ in \cref{eq:ImA_resoances} converges but will diverge at $s> \lambda^2$ and must be re-summed into the Regge-behaved contributions of the previous subsection. In this way, the momentum scale $\Lambda$ mimics the radius of interaction~\cite{Gribov:2009cfk}.
Because each isobar is an infinite sum of simple one-particle exchanges, \cref{eq:gen_isobar_decomp} is a crossing symmetric sum of Reggeons each realized by a van-Hove-like model~\cite{vanHove:1967zz}. Furthermore, $\Lambda$ also marks the divergence of the crossed channel PW series and therefore plays the role of the semi-major axis of the (small) Lehmann ellipse~\cite{Lehmann:1958ita}. 

As shown, the amplitude given by \cref{eq:gen_isobar_decomp,eq:mF_pm,eq:f(alpha_z)} corresponds to the sum of three terms each of which 
is responsible for the resonances and the Regge behavior in a specific channel. This deviates from the usually held notions of duality, which argue instead for a decomposition in which the Regge behavior in one channel is dual to the resonances in another. More general definitions, such as proposed by Ref.~\cite{Childers:1969am}, i.e., that duality only requires resonances and Reggeons to emerge from the same function $\as$, are still satisfied.
A more detailed discussion and comparison with dual models is provided in \cref{app:duality}.

%%%%%%%%%%%%%%%%%%%%%%%%%%%%%%%%%
\subsection{The Pomeron isobar}
\label{sec:pomeron}
%%%%%%%%%%%%%%%%%%%%%%%%%%%%%%%%%

Before concluding the discussion on the general properties of our hypergeometric isobars, we consider how the Pomeron trajectory may be incorporated in our model.
The Pomeron shares the quantum numbers of the vacuum and is typically ascribed to the exchange of gluonic degrees of freedom~\cite{Gell-Mann:1964bha}. The Pomeron is phenomenologically well established and required to describe diffractive peaks in a wide array of hadronic scattering at high energies (cf.\ Ref.~\cite{Donnachie:2002en} and references therein). In the low-energy regime, these gluonic exchanges could correspond to glueballs, i.e., resonances made entirely of gluons. Although the existence of glueballs has long been conjectured to lie on the Pomeron trajectory~\cite{Mathieu:2008me,Ochs:2013gi,Szanyi:2019kkn,Llanes-Estrada:2021evz}, no state has been unambiguously identified by experiment. Alternative interpretations of the Pomeron trajectory emerging from purely non-resonant exchanges have also been proposed~\cite{Freund:1967hw,Harari:1968jw}.

As such, we construct a special non-resonant isobar, which allows the Pomeron trajectory $\aspom$ to contribute to the Regge behavior at high energies without adding direct-channel poles at low energies.\footnote{We will only discuss the leading trajectory and do not discard the possibility of glueballs appearing from daughter trajectories.} In practice, this includes making the assumption that the Pomeron is a simple Regge pole. While this is a phenomenologically reasonable assumption, the true nature of the Pomeron with respect to the complex $j$-plane may be more complicated~\cite{Badatya:1984sp,Donnachie:2002en}.

We define the non-resonant isobar as 
    \begin{align}
        \label{eq:f(alpha_z)_Pomeron}
        \fasNR &= \frac{\gaussFR{1, - \aspom}{ 1 - \aspom}{\hat{\nu}_s}}{\aspom-1}
        ~.
    \end{align}
This is \cref{eq:f(alpha_z)} with $\jmin = 0$ and the $\Gamma(-\as)$ prefactor replaced by a single pole at $\aspom=1$.
Through the direct calculation of the Regge limit, this will yield the Regge behavior (cf.\ \cref{eq:regge_behavior}),
    \begin{equation}
        \label{eq:pomeron_regge}
        \mathbb{P}(s,t) = \frac{\beta_\pom(t) \, \xi_+(t)}{\atpom -1} \, \left(\frac{s}{2\Lambda^2} \right)^{\alpha_\pom(t)} ~,
    \end{equation}
with $\beta_\pom(t) = g^2_\pom \, \Gamma(1+\atpom)$. Since the Pomeron has positive signature and $\alpha_\pom(0) \sim 1$, the trajectory can be expanded around the forward peak at $t=0$
    \begin{equation}
        \label{eq:pomeron_forward}
        \mathbb{P}(s,t\to 0) = \frac{i \pi}{2} \, g^2_\pom \times \left(\frac{s}{2\Lambda^2}\right) ~,
    \end{equation}
which is entirely imaginary as expected from high-energy phenomenology. Because the pole at $\aspom=1$ occurs at a wrong-signature value of $j$, it is canceled by a zero of the signature factor $\xi_+(t)$ to give a finite contribution at forward $t$.

Checking the behavior of the other isobars in the Regge limit, it is straightforward to see the usual assumption of $\aupom\to-\infty$ as $u\to-\infty$ retains vanishing behavior faster than  \cref{eq:t_limit_lm1}.
Similar to \cref{eq:s_limit_plus}, ensuring the $s$-channel isobar vanishes entails a requirement on the real and imaginary parts of $\aspom$. Following the analogous derivation to obtain \cref{eq:Imalpha_bound}, one can show that if $|\aspom| \leq s^{1-\epsilon}$ is bounded above for some $0 \leq \epsilon < 1$ then the $s$-channel isobar will be exponentially suppressed if asymptotically  
    \begin{equation}
        \label{eq:pomeron_bound}
        \Im \aspom > \left(\frac{4-2\epsilon}{\pi}\right)\, \Re\aspom \, \log \left(\frac{s}{4\Lambda^2}\right) ~.
    \end{equation}
In the resonance region, we expand in $\nu_s$ and write the expansion
    \begin{align}
        \label{eq:fas_Pom_resonances}
        \fasNR =  \frac{\aspom}{\Gamma(2-\aspom)} \sum_{j=0}^\infty \, \frac{(\qhatzs)^j}{j-\aspom} ~.
    \end{align}
At $\aspom = 0$, the pole in the first term of the sum is canceled by the numerator prefactor. Similarly, for all integer $\aspom = j\geq 2$, the poles are canceled by the $\Gamma$ factor in the denominator. Thus, the only pole contained in \cref{eq:fas_Pom_resonances} arises from the $j=1$ term, which will be removed by the parity factor $\tau_j$ in the positive signature combination $\mF{+}(s,z_s)$, cf.\ \cref{eq:mF_pm,eq:mF_resonance}.\footnote{Note that an arbitrary number of poles at different \textit{odd} values of $j$ can be introduced and will be canceled by the signature factor. Thus, \cref{eq:pomeron_regge} is a minimal choice of $t$-dependence for the Regge residue.} The isobar \cref{eq:f(alpha_z)_Pomeron} therefore introduces no poles and is consistent with phenomenological expectations of the Pomeron.

%%%%%%%%%%%%%%%%%%%%%%%%%%%%%%%%%
\section{Dispersive trajectories}
\label{sec:trajectories}
%%%%%%%%%%%%%%%%%%%%%%%%%%%%%%%%%

As we have demonstrated, \cref{eq:mF_pm,eq:f(alpha_z)} will recover many appealing features of amplitudes in both the resonance and asymptotic regimes if certain assumptions on the RTs are made. The isobar model is ultimately ineffective, however, unless RTs can be constructed to satisfy all requirements while remaining flexible enough to fit scattering data. In this section, we explore plausible models for this purpose.

The assumption that $\as$ is an analytic function with only a RHC and bounded above by $s$ means it can be written as a once-subtracted dispersion relation 
    \begin{equation}
        \label{eq:disp_one_sub}
        \alpha(s) = \alpha(0) + \frac{s}{\pi}\int_{4m^2}^\infty \dd s' \frac{\Im\alpha(s')}{s' \, (s' - s - \ieps)} ~.
    \end{equation}
Dispersion relations have long been a starting point for constructing Regge trajectories as they provide an effective way to incorporate dynamical input while preserving analyticity~\cite{Mandelstam:1968zza,Epstein:1968vaa,Chu:1968ctr,Atkinson:1969fe,Degasperis:1970us,Fiore:2000fp,Londergan:2013dza,JointPhysicsAnalysisCenter:2024znj}. We disallow a linear term, i.e., a second subtraction, not only because it would violate previously mentioned asymptotic bounds, but it would mean the slope of $\Re\as$ and therefore the particle spectrum is determined by parameters external to the reaction dynamics~\cite{Collins:1968akw}. We must thus reconcile the phenomenologically observed linearity of $\Re\as$ for many mesons with a RT given by the form in \cref{eq:disp_one_sub}.

As seen in \cref{eq:ImA_resoances}, unitarity must be implemented in our model through $\Im\as$ and can thus help guide the functional form of the imaginary part. Taking the $j$-th PW projection of \cref{eq:ImA_resoances} gives
    \begin{align}
        \label{eq:imAj}
        \Im \, &\PW_j(s) 
        \\ 
        &=\sum_i  g^2_i \, n_{ij} \, \tau_j \,  \mu_{jj} \, \qhats^{2j} \, \frac{\Im \alpha_i(s)}{|j - \alpha_i(s)|^2} \,    + \O{\qhats^{2(j+2)}}
        \nonumber~,
    \end{align}
where all contributions of poles at $\as > j$ appear with higher powers of momentum.
Approaching threshold, these higher terms vanish faster than the leading pole, and the imaginary part of the full amplitude will be dominated by \cref{eq:imAj} with $j$ replaced with the lowest physical spin. Similarly, since the first pole in each isobar term, \cref{eq:fas_poles}, occurs at $\as = \jmin$, the lowest PW to which each $\as$ will contribute also has $j=\jmin$.\footnote{Because of the parity factor $\tau_j$, $\jmin$ can technically be selected as a wrong-signature value and the sum will effectively start at $j = \jmin+1$. For simplicity, here we assume that $\jmin$ and $j$ have the same parity.} Thus, examining \cref{eq:imAj} together with the unitarity condition of PWs~\cite{Collins:1977jy}
    \begin{equation}
        \label{eq:PW_unitarity}
        \Im \PW_j(s) = \rho(s) \, |\PW_j(s)|^2 ~,
    \end{equation}
where $\rho(s) = 2 q_s/\sqrt{s}$ is the relativistic two-body phase space, 
each RT must satisfy
    \begin{equation}
        \label{eq:imAlpha_sth}
        \Im\alpha_i(s \to 4m^2) \propto \rho(s) \,\qhats^{2\jmin} \, g^2_i 
        ~,   
    \end{equation}
for the lowest PWs to fulfill unitarity at threshold.
Note that even if there are multiple poles or crossed channel contributions, \cref{eq:imAlpha_sth} must still hold to ensure each pole term has the required powers of momentum.  
Although not considered here, in a coupled-channel scenario similar limits could be derived for each multi-particle threshold. 

In addition to constraining the behavior near threshold, analyticity and unitarity principles can be used to constrain the asymptotic behavior. Along fairly general arguments, for instance, the asymptotic growth of any complex trajectory should be bounded by a square root up to possibly arbitrary $\log s$ factors~\cite{Childers:1968vnm}. Combined with \cref{eq:Imalpha_bound} to ensure polynomial boundedness~\cite{Bugrij:1973ph,Trushevsky:1975yf}, however, this bound becomes stricter with
    \begin{equation}
        \label{eq:childers_bound}
        |\alpha(s\to\infty)| \leq \sqrt{s} \, \log s ~.
    \end{equation}
Because \cref{eq:disp_one_sub} is defined through a once-subtracted dispersion relation, the positivity and unboundedness of $\Im\alpha(s\geq 4m^2)$ will guarantee $\alpha(s\to-\infty) \to -\infty$~\cite{Gribov:1962gb}.

The construction of RTs that satisfy conditions like \cref{eq:disp_one_sub,eq:imAlpha_sth,eq:childers_bound} is not entirely new. For example, Ref.~\cite{Fiore:2000fp} constructs a model for the imaginary part of $\alpha(s)$ built up from terms of the form:
    \begin{equation}
        \label{eq:Fiore_ImAlpha}
            \Im\alpha(s) = \sum_{i} g^2_i \, \sqrt{s - s_i} \, \left(\frac{s-s_i}{s}\right)^{\lambda_i}\theta(s-s_i) ~.
    \end{equation}
Here the sum refers to the openings of multi-particle thresholds located at $s_i$. The exponents $\lambda_i$ determine the vanishing behavior at each threshold and are used to enforce unitarity constraints, e.g., by taking $\lambda_i = \Re\alpha(s_i)$ or, in the case of \cref{eq:imAlpha_sth}, choosing the lowest threshold to satisfy  $\lambda_0 = \jmin$.

Because \cref{eq:Fiore_ImAlpha} behaves as $\sqrt{s}$ asymptotically, the real part of the trajectory from evaluating \cref{eq:disp_one_sub} will asymptotically approach a constant for $s$ greater than the highest considered threshold. Therefore, any trajectory of this type will trivially satisfy the bounds in \cref{eq:Imalpha_bound,eq:pomeron_bound}. 
Achieving the quasi-linear behavior observed in the particle mass spectra, however, requires adding multiple higher thresholds. In addition, while the form \cref{eq:Fiore_ImAlpha} can reproduce the branch point structure required by unitarity and analyticity, the precise $s$ behavior is fairly rigid and does not allow much flexibility to unitarize PWs of the form \cref{eq:imAj}.
Thus, we seek a different functional form with which to describe resonances in conjunction with the isobar model in \cref{sec:hypergeo_iso}. 

We will parameterize the RTs with a logarithmic form:
    \begin{equation}
        \label{eq:log_form}
        \Im\alpha(s) = \frac{\gamma}{\pi} \log\left(1 + \frac{\pi}{\gamma} \, \rho(s)  \, r(s) \right) \theta(s-4m^2)~,
    \end{equation}
with a constant $\gamma > 0$ and a real function $r(s)$. At \mbox{$s\to 4m^2$} and $\rho(s) \to 0$ we may expand the logarithm,
    \begin{equation}
        \label{eq:log_sth}
        \Im\alpha(s \to 4 m^2) = \rho(s) \,r(s) ~,  
    \end{equation}
which is independent of $\gamma$. The function $r(s)$ is assumed to be free of singularities along real $s \geq 4m^2$ but is otherwise completely general. It can thus be used to help us enforce unitarity constraints.
For a single isolated pole with $j=\jmin$, for instance, \cref{eq:log_sth,eq:imAlpha_sth} would imply $r(s) = g^2 \, \mu_{\jmin\jmin} \, \qhats^{2\jmin} +  \O{\qhats^{2\jmin +1}}$, where the coupling $g$ and $\qhats^2$ should be the same as those appearing in the isobars. We will thus assume that $r(s)$ can be written as:
    \begin{equation}
        \label{eq:rs}
        r(s) = \sum_{k=0}^{N} \, c_k \, \qhats^{2k}  + c_{\alpha} \, \qhats^{2\Re\alpha(s)} ~.
    \end{equation}
This is a general, but still fairly minimal, parameterization of the possible function $r(s)$. For example, \cref{eq:rs} can be multiplied by any overall power of $s$ or $\log s$, but this is omitted for simplicity.
The first term in \cref{eq:rs} allows energy behavior bounded by an arbitrary finite-order polynomial of $\qhats^2$, or equivalently of $s$, with real coefficients $c_k$. The second term, on the other hand, enforces a Regge-like power-law behavior. The overall logarithm in \cref{eq:log_form} means any fixed power behavior in $s$ can be added without having exponential growth and thus the first term allows the trajectory to be flexible enough to parameterize amplitudes at finite $s$ when expanded for $\qhats^2 < 1$, e.g., through \cref{eq:log_sth}. Individual $c$'s in \cref{eq:rs} can be negative but for $\Im\alpha(s)$ to be real and positive on the real axis requires $r(s) > 0$ for all $s\geq 4m^2$.

Taking the limit $s\to \infty$ of \cref{eq:log_form} with \cref{eq:rs}, 
we see:
    \begin{align}
        \label{eq:s_ininity_log}
        \Im\alpha&(s\to \infty) =\frac{\gamma}{\pi} \,  \log \left( c_{N} \, \qhats^{2N} + c_\alpha \, \qhats^{2\Re\alpha(s)} \right) 
        ~,
    \end{align}
where the asymptotic behavior is dictated by the behavior of the last term. Because $\Re\alpha(s)$ is assumed to be unbounded, we have two possibilities depending on $\Re\alpha(s) \to \pm\infty$. If $\Re\alpha(s) \to -\infty$ the Regge-like term will quickly vanish as $\qhats >1$ and we have $\Im\alpha(s\to\infty) = (\gamma/\pi) \, N \, \log (s/4\Lambda^2)$. If instead $\Re\as \to +\infty$, then $\Im\alpha(s\to\infty) = (\gamma/\pi) \Re\as \, \log(s/4\Lambda^2)$ and the bound \cref{eq:Imalpha_bound} can be encoded by requiring $\gamma >1$. 
The trade-off, however, is that when inserted in \cref{eq:disp_one_sub}, $\as$ is now defined through a non-linear integral equation, which must be solved numerically. 

Although it is not obvious, this integral equation admits stable solutions, which satisfy $\Re\alpha(s) \sim \sqrt{s}$ and $\Im\alpha(s) \sim \sqrt{s} \, \log s$ asymptotically and thus saturate the bound in \cref{eq:childers_bound}. To illustrate this point, we will solve the integral equation by fitting the masses of particles on the exchange degenerate $\rho$ -- $a_2$ RT. This is intended as a proof-of-concept and the $\rho$ trajectory will be revisited in \cref{sec:I=1} when considering $\pi\pi$ scattering using more in-depth unitarity constraints combined with the isobar model of \cref{eq:f(alpha_z)}.  

Analogous to the analysis in Ref.~\cite{Fiore:2000fp}, we adopt an iterative fitting procedure, where we start with an initial guess for $\Re\alpha(s)$ and fix free parameters by fitting \cref{eq:disp_one_sub} with a least squares minimization:
    \begin{equation}
        \label{eq:leastsquares_spectra}
        d^2 =  \sum_i \, \left[ \left(\Re\alpha(m_i^2) - j_i \right)^2 +  \left(\Gamma(m_i^2) - \Gamma_i \right)^2\right] ~,
    \end{equation}
where $m_i$, $\Gamma_i$, and $j_i$ are the masses, widths, and spins of the $\rho$/$a_2$ mesons and their orbital excitations. To connect the width with the RT we expand the width function for narrow resonances given by~\cite{Gribov:2003nw}:
    \begin{equation}
        \Gamma(s) = \frac{\Im\alpha(s)}{\sqrt{s} \, \Re \alpha^\prime (s)} ~.
        \label{eq:BW_width}
    \end{equation}
As our primary interest is in the existence and properties of a solution and not the numerical values of parameters, we will ignore any errors associated with the input masses and widths.
After a good fit is found, the initial guess of $\Re\as$ is updated with an interpolation of the previous best-fit real part and the trajectory is fit again. This procedure is continued until a stable solution is found.

We consider the masses and widths of isovectors of both signatures up to $j\leq 6$ from the Review of Particle Physics (RPP)~\cite{ParticleDataGroup:2022pth}. In the absence of individual PWs, only the last term in \cref{eq:rs} is kept, i.e., all $c_k=0$, and fix $\alpha(0) = 0.5$ and $\Lambda = \sqrt{2}\GeV$ for simplicity. This latter value is chosen such that $\sqrt{s} = \lambda \approx \sqrt{2}\Lambda \approx 2\GeV$ coincides with the energy at which Regge behavior appears to begin in the $\pi\pi$ total cross section~\cite{Pelaez:2003ky,Caprini:2011ky,Halzen:2011xc}. In all numerics, we consider units of GeV unless otherwise stated and start with the initial guess $\Re\alpha(s) = (0.5 + 0.9 \, s)/\sqrt{1+s/20}$
and fit the remaining two parameters $\gamma$ and $c_\alpha$. A reasonably stable solution is found after about four iterations of the integral equation as seen in the best-fit parameters tabulated in \cref{tab:spectrum_fit}. 
    \begin{table}[t]
        \centering
        \caption{Results for the iterative fitting procedure for the exchange degenerate $\rho$ -- $a_2$ trajectory. The fit value at the $i$-th fit iteration is shown for the two free parameters $\gamma$ and $c_\alpha$. }
        \label{tab:spectrum_fit}
        \begin{ruledtabular}
        \begin{tabular}{ccc}
            $i$ & $\gamma$ & $c_\alpha$ \\
            \hline
             0 & 1.345 & 4.953 \\
             1 & 1.101 & 3.100 \\
             2 & 1.072 & 3.612 \\
             3 & 1.083 & 3.565 \\
             4 & 1.085 & 3.566 \\
             5 & 1.082 & 3.573 \\ 
             \hline
             10 & 1.082 & 3.571 \\
             15 & 1.083 & 3.570 \\
             20 & 1.083 & 3.569 \\
        \end{tabular}
        \end{ruledtabular}
    \end{table}

\begin{figure*}[t]
        \centering
        \includegraphics[width=0.49\textwidth]{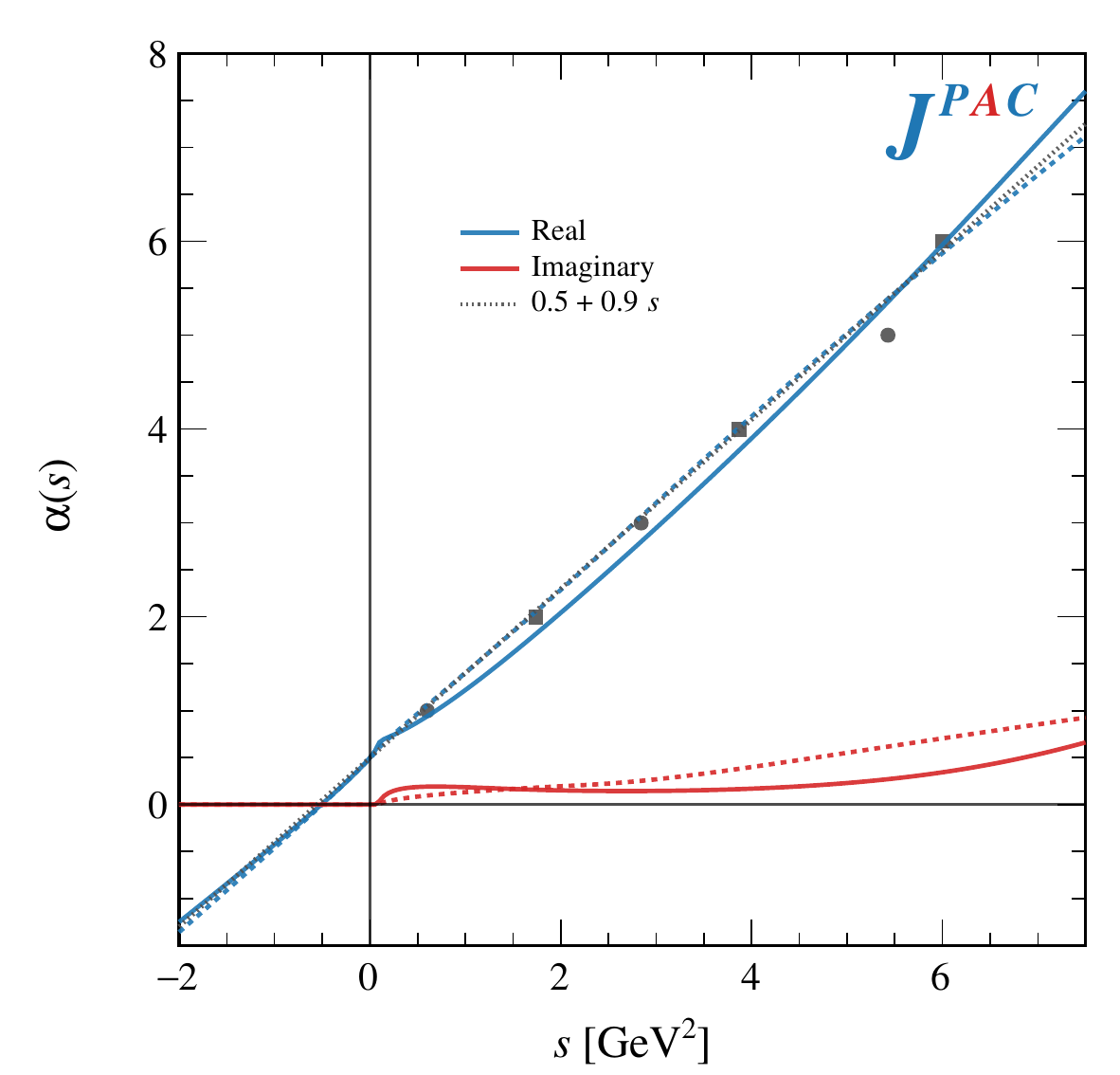}
        \includegraphics[width=0.49\textwidth]{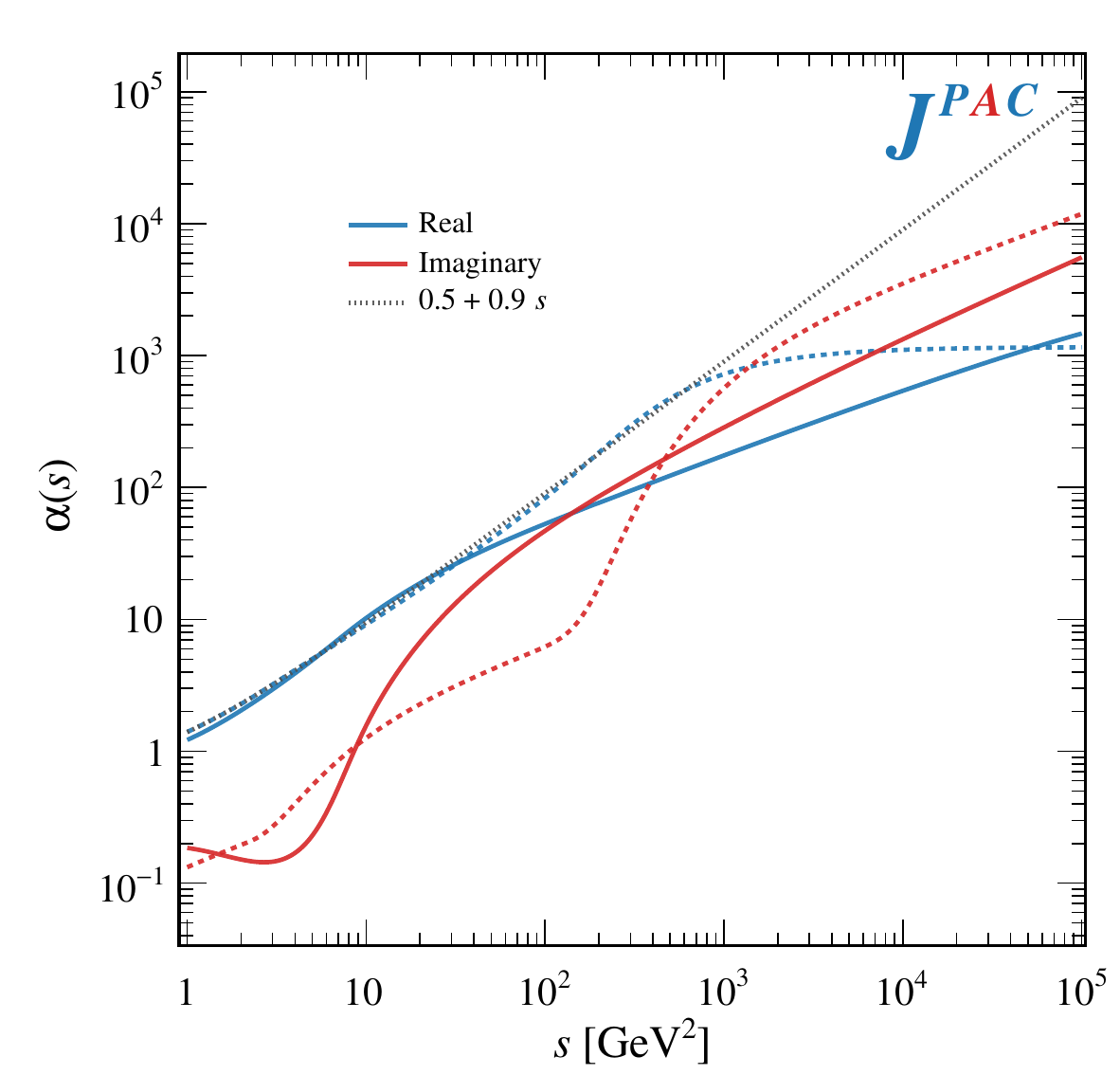}
        \caption{Results for the exchange degenerate $\rho$ -- $a_2$ trajectory in the resonance region (left) and at asymptotic energies (right). The results using \cref{eq:log_form,eq:rs} (solid) are compared to the model of Ref.~\cite{Fiore:2000fp} using \cref{eq:Fiore_ImAlpha} (dashed) as well as the canonical linear $\rho$ trajectory (black dotted).}
        \label{fig:spectrum_fit}
    \end{figure*}

In \cref{fig:spectrum_fit} the resulting RT after 20 iterations is plotted compared to both the phenomenological linear trajectory \mbox{$(0.5 + 0.9 \, s)$}~\cite{Collins:1977jy} and the results of Ref.~\cite{Fiore:2000fp}. 
In this comparison, we note several things:
first, both \cref{eq:Fiore_ImAlpha,eq:log_form} achieve approximate linearity in the resonance region, however, this is accomplished by completely different mechanisms. For \cref{eq:Fiore_ImAlpha}, effective multi-body threshold openings are required to ``kick up" the imaginary part and prevent the real part from saturating to a constant. The linearity of Regge trajectories is therefore assumed to be an inherently inelastic phenomenon. \Cref{eq:log_form}, however, accomplishes the quasi-linear behavior using only a single threshold. Although we do not explore this here, considering additional thresholds can modify the slope of the trajectory and possibly lead to an asymptotically constant real part. In this way, it is actually the termination of resonances which is a multi-threshold effect. Since the termination of the infinite tower of excited resonances is proposed to be related to screening from coupled channels~\cite{Brisudova:1998wq,Kholodkov:1991hx}, the interpretation of the latter mechanism seems more plausible. 

Second, $\Re\alpha(s)$ continues to grow indefinitely, but slows from approximately linear to square-root behaved. A model with infinitely many poles, such as in \cref{eq:fas_poles}, will thus indeed have infinitely many resonances appearing as orbital excitations. Note that, unlike narrow resonance models, $\Im\alpha(s)$ also grows indefinitely and at a faster rate. This has the effect of moving higher-$j$ poles deeper and deeper into the complex plane, such that the infinite tower of resonances is indiscernible from a non-resonant background above some energy scale. 

Finally, we note that the best-fit value of $\gamma \approx 1.08 > 1$ means the asymptotic behavior of $\alpha(s)$ satisfies the bounds in \cref{eq:childers_bound,eq:Imalpha_bound}. A crossing symmetric combination of isobar terms of the form in \cref{eq:mF_pm} with the trajectory as in \cref{fig:spectrum_fit} will thus be properly Regge-behaved at high energies.

%%%%%%%%%%%%%%%%%%%%%%%%%%%%%%%%%
\section{Application to \texorpdfstring{$\pi\pi$}{pion-pion} scattering}
\label{sec:application}
%%%%%%%%%%%%%%%%%%%%%%%%%%%%%%%%%

As we have demonstrated, the isobar model constructed in \cref{sec:hypergeo_iso} offers a unified description of low-energy resonances and high-energy Regge behavior through analyticity in both energy and angular momentum. The key ingredients to accomplish this are RTs, which encode all the relevant dynamical information. In \cref{sec:trajectories}, an RT model that satisfies all the requirements to realize the isobar model in \cref{eq:gen_isobar_decomp} is discussed. In this section we now combine the results of the previous sections to consider $\pi\pi$ scattering and study the $\rho$ and $\sigma/f_0(500)$ resonances as a benchmark of the presented theoretical framework using as few parameters as possible. This is not intended to be a precision study of these resonances.

Because the pion is an isovector, \cref{eq:gen_isobar_decomp} must be generalized to accommodate the scattering of different isospin states. Since the isobars in \cref{eq:mF_pm} already have definite signature, such a generalization is trivially accomplished by defining isobars with a definite isospin $I = 0,1,2$ as:
    \begin{align}
        \label{eq:mF_I}
        \mF{I}&(s,z_s) \\
        &=\sum_i \frac{g_{i}^{2}}{2}\left[F(\alpha^{I}_i(s), \nu_s) + (-1)^I \, F(\alpha^{I}_i(s),-\nu_s)\right] ~. \nonumber
    \end{align}
This isobar transforms as $\mF{I}(s,z_s) = (-1)^I \, \mF{I}(s,-z_s)$, as required by Bose symmetry. We assume that the RTs, which are summed over, also carry a definite isospin. While exchange degeneracy, i.e., the approximate equality of RTs with similar quantum numbers~\cite{Desgrolard:2000sf}, can still be imposed, we will not require it. In general, each trajectory is only responsible for generating the resonances of a single signature and isospin. 

The analog of \cref{eq:gen_isobar_decomp} for the $s$-channel isospin amplitudes can be written by constructing a crossing symmetric combination of terms given by \cref{eq:mF_I}~\cite{Albaladejo:2018gif,Albaladejo:2019huw}
    \begin{align}
        \label{eq:mA_I}
        \Amp^I(&s,t,u) = \mF{I}(s, z_s) \\
        &+ \sum_{\Ip} C^{I\Ip}_{st} \left[ \mF{\Ip}(t,z_t) + (-1)^{I+\Ip} \, \mF{\Ip}(u,z_u) \right] ~,
        \nonumber
    \end{align}
where $C^{I\Ip}_{st}$ are the elements of the isospin crossing matrix~\cite{Chew:1960iv,Chew:1961yz}
    \begin{align}   
        \label{eq:Cmatrix}
        C_{st} =
        \setlength\arraycolsep{4pt}
        \def\arraystretch{1.5}
        \begin{pmatrix}
        \frac{1}{3}&1&\frac{5}{3}\\
        \frac{1}{3}&\frac{1}{2}&-\frac{5}{6}\\
        \frac{1}{3}&-\frac{1}{2}&\frac{1}{6}\\
        \end{pmatrix}~.
    \end{align}
An explicit demonstration that \cref{eq:mA_I} is crossing symmetric, i.e., the decomposition with respect to isospin defined in the $t$- or $u$-channels is identical, is relegated to \cref{app:isospin}.

Since each isobar will only have poles in its energy variable, \cref{eq:mA_I} will only have $s$-channel resonances of isospin $I$ coming from the first term. The remaining terms thus represent the exchange of all-isospin particles in the $t$- and $u$-channels. Assuming well-behaved RTs in each channel, taking the $s \to \infty$ limit with fixed $t \lesssim 0$, \cref{eq:mA_I} yields the Regge-behavior
    \begin{equation}    
        \label{eq:mA_I_regge}
        \Amp^I(s,t,u) \to \sum_\Ip \, C^{I\Ip}_{st} \, \left[ \sum_{i} \,\mathbb{R}_i^\Ip(s,t) \right]~,
    \end{equation}
with each $\mathbb{R}^\Ip_i$ given by \cref{eq:regge_behavior} with respect to $\alpha_i^\Ip(t)$. Clearly, \cref{eq:mA_I_regge} represents the exchange of Reggeons of all isospins in the $t$-channel with the correct coefficients from crossing as expected from high-energy $\pi\pi$ scattering~\cite{Ananthanarayan:2000ht,Pelaez:2003ky,Caprini:2011ky}.

Although the exploratory $\rho$ trajectory in \cref{fig:spectrum_fit} reasonably reproduces the resonance region with only a single threshold, this result should be interpreted with caution. Beyond the mass of the lowest-lying $\rho$, inelastic thresholds become increasingly important, e.g., the $\rho_3(1690)$ decays primarily into $4\pi$. Constraining a RT in a broad range of energies is thus inherently a coupled-channel problem. For this first study, then, we will restrict ourselves to the $\pi\pi$ scattering in the isospin limit below the $K\bar{K}$ threshold.  In these energies, the primary contributions come from two-body dynamics and the RTs contain only one relevant branch point~\cite{Collins:1977jy}. The amplitude can thus be effectively constrained with elastic unitarity in order to benchmark the extraction of the RTs of mesons in this mass region. 

Fixing $\Lambda = \sqrt{2}\GeV$ (and therefore $\lambda \simeq \sqrt{2} \, \Lambda = 2 \GeV$) at the observed scale of Regge physics in $\pi\pi$ scattering as before, the elastic region lies well within the boundary $s<\lambda^2$ where our amplitude is a genuine isobar model as discussed in \cref{sec:resonances}. We can thus take the $j$-th PW projection, which is decomposed into separate contributions from direct and crossed channel isobars:
    \begin{equation}
        \label{eq:A_Ij}
        \PW^I_j(s) = f_j^I(s) + \sum_{\Ip} \, C_{st}^{I\Ip} \, \tilde{f}_j^\Ip(s) ~.
    \end{equation}
The direct channel term contains the RHC and is given by the projection of all the $s$-channel poles, the projection of which can be written explicitly using \cref{eq:mF_resonance}
    \begin{equation}
        \label{eq:f}
        f_j^I(s) = \tau_{I+j}\, \sum_i \,  \sum_{k=0}^\infty   \qhats^{2(j+2k)} \, \left[\frac{g_i^2 \, n_{i,j+2k} \, \mu_{j+2k,j}}{j-(\alpha^I_i(s)-2k)}\right] ~.
    \end{equation}
The ``inhomogenous" term on the other hand is given by the projection of the crossed channel isobars:
    \begin{equation}
        \label{eq:f_tilde}
        \tilde{f}^I_j(s) = \int_{-1}^1 \dd z_s \, P_j(z_s) \, \mF{I}(t(s,z_s),z_t(s,z_s))  ~,
    \end{equation}
which will generate the LHC of $\PW_j^I(s)$ and cannot be done in closed form. The structure of the PW in \cref{eq:A_Ij} is intentionally written to mirror the structure of the KT decomposition~\cite{Khuri:1960zz,Albaladejo:2019huw,Stamen:2022eda}. 

Since imposing unitarity in the KT formalism requires solving systems of coupled integral equations for each isospin simultaneously, they are typically very challenging. In our formalism, the analogous integral equations will be non-linear. Because of this, we first demonstrate that unitarity can be imposed by solving the ``homogeneous" equation 
    \begin{equation}
        \label{eq:KT_homogeneous}
        \Im \PW_j^I(s) = \Im f_j^I(s) = \rho(s) \, |f_j^I(s)|^2 ~,
    \end{equation}
which ignores the contribution from the crossed channels in the second term of \cref{eq:A_Ij}.
In the conventional KT formalism, this reduces to a Muskhelishvili--Omn\`es problem and is readily solved in terms of the scattering phase shift~\cite{Muskhelishvili:1958,Omnes:1958hv}. In the language of \cref{eq:A_Ij}, on the other hand, the homogeneous solutions will decouple the different isospins and yield RTs without corrections from final-state interactions. 
Thus, despite not involving the full crossing symmetric model in \cref{eq:mA_I}, solving the homogeneous problem is a highly non-trivial and necessary first step towards a full ``KT with Regge poles" analysis to be done in the future.\footnote{We have chosen to implement unitarity at the level of individual PWs, but because our isobars incorporate the infinite tower of increasing spin, one could, in principle, try to unitarize the full amplitude.} 

As described in \cref{sec:resonances}, the imaginary part of the amplitude arises from the imaginary part of the RTs. Using \cref{eq:log_form}, the degrees of freedom with which to incorporate unitarity are the coefficients of the  $\qhats^2$ polynomial contained within \cref{eq:rs}. In practice, because this polynomial is of fixed order, unitarity can only be imposed up to a certain momentum scale corresponding to the first power of $\qhats^2$ which is not considered.
Luckily, with the scale parameter $\Lambda = \sqrt{2}\GeV$, the region below the $K\bar{K}$ threshold, i.e., with $s\lesssim 1\GeV^2$, has $\hat{q}^2_{s} \lesssim 0.12$ and the sums over powers of momentum in \cref{eq:f} converge very quickly. Elastic unitarity can thus be implemented numerically with only a few terms. 

As our primary focus is the application to hadron spectroscopy, we will focus on the $I=0$ and $1$ channels. Resonances with $I=2$ would correspond to doubly-charged mesons and are not observed in nature. From \cref{eq:ImA_resoances}, we see that PWs in our formalism can only achieve a non-zero imaginary part with an explicit RT in the direct channel and thus parameterizing any $I=2$ amplitude would indeed require constructing at least one $\alpha^{I=2}(s)$. Luckily, exotic mesons can be avoided with a RT that never crosses positive even integers or through a non-resonant isobar analogous to that in \cref{sec:pomeron}. We, however, do not pursue this further.

%%%%%%%%%%%%%%%%%%%%%%%%%%%%%%%%%
\subsection{\texorpdfstring{$I=1$}{I=1} and \texorpdfstring{$\alpha_\rho(s)$}{alpha rho(s)}}
\label{sec:I=1}
%%%%%%%%%%%%%%%%%%%%%%%%%%%%%%%%%
The isospin-1 channel of the elastic $\pi\pi$ spectrum is well known to be dominated by the $\rho$ resonance in the $P$-wave. As such, we include only a single trajectory $\alpha_\rho(s) \equiv \alpha^{I=1}(s)$ with $\jmin = 1$, such that the isobar \cref{eq:mF_I} takes the form 
    \begin{equation}
        \label{eq:F_1}
        \mF{1}(s,z_s) = \frac{g^2_\rho}{2} \left[ F(\alpha_\rho(s), \,\nu_s) -  F(\alpha_\rho(s), \,- \nu_s)\right]~.
    \end{equation}
We use a trajectory given by \cref{eq:disp_one_sub,eq:log_form} with \cref{eq:rs} containing two terms:
    \begin{equation}
        \label{eq:r_rho}
        r_\rho(s) = \frac{g_\rho^2}{3} \, \qhats^2 + c_\rho \, \qhats^{2(1+\Re\alpha_\rho(s))} ~,
    \end{equation}
in order to unitarize up to order $\O{\qhats^2}$, which encompasses only the $P$-wave (i.e., the $F$-wave is $\O{\qhats^{6}}$). Expanding the logarithm around small $\qhats^2$ with \cref{eq:log_sth}, the $P$-wave projection of \cref{eq:F_1} is given by 
    \begin{equation}
        \label{eq:im_pw_rho}
        \Im f_1^1(s) =  \rho(s) \left|f_1^1(s)\right|^2 +\O{\qhats^{2(2+\Re\alpha_\rho(s))}} ~.
    \end{equation}
Since $\Re\alpha_\rho(s) \gtrsim 0.5$ is expected above threshold, e.g., similar to \cref{fig:spectrum_fit}, the second term will always be subleading and \cref{eq:im_pw_rho} reproduces \cref{eq:KT_homogeneous} to leading order.

The first higher-order term in \cref{eq:im_pw_rho} scales with $\Re\alpha_\rho(s)$ and will be the dominant correction at small and intermediate $s$. However, as $s$ increases, and $\Re\alpha_\rho(s) > 2$, the next-to-leading order term will instead be the next fixed power of $\qhats^2$ in \cref{eq:f}. The constant $c_\rho$ in \cref{eq:r_rho} is thus left as a free parameter to incorporate the small contributions from these terms in the unitarity equation at low energies. We fix $\alpha_\rho(0) = 0.491$ to the value extracted from charge-exchange $\pi N$ scattering in Ref.~\cite{Barnes:1976ek} and thus have three parameters left to be determined: $g^2_\rho$, $\gamma_\rho$, and $c_\rho$.  We adopt an iterative fitting procedure as in \cref{sec:trajectories} to simultaneously solve the integral equation for $\alpha_\rho(s)$ as well as fix parameters by comparing them to data. Using the same initial guess as before, we now minimize the distance squared of the resulting $P$-wave projection of \cref{eq:F_1} to known $\pi\pi$ PWs 
    \begin{equation}
        \label{eq:leastsquares}
        (d^2)^I_j = \sum_i \left|f_j^I(s_i) - (t_\text{GKPY})_j^I(s_i) \right|^2 ~.
    \end{equation}
Here $(t_\text{GKPY})^I_j$ is the central value of PWs with a given isospin as determined by the Madrid group~\cite{Garcia-Martin:2011iqs}. Once again, as an exploratory study, we ignore the errors associated with these PWs. By fitting both real and imaginary parts of the PWs simultaneously, the constraint of elastic unitarity is incorporated into the RT. For the present case of the $\rho$ trajectory, $f_1^1(s_i)$ is given by projecting \cref{eq:F_1} onto the $P$-wave and we choose 10 evenly spaces points between $4m_\pi^2$ and $1\GeV^2$ as the sampled energies $s_i$.

    \begin{table}[t]
        \centering
        \caption{Results for the iterative fitting procedure of the $\rho$ trajectory using low-energy unitarity.}
        \label{tab:rho_pars}
        \begin{ruledtabular}
        \begin{tabular}{cccc}
            $i$ & $g_\rho^2$ & $\gamma_\rho$ & $c_\rho$ \\
            \hline
             0  & 4.269 & 1.302 &  2.818 \\
             1  & 4.262 & 1.074 &  3.231 \\
             2  & 4.263 & 1.071 &  3.241 \\
             3  & 4.263 & 1.079 &  3.142 \\
             4  & 4.263 & 1.080 &  3.212 \\
             5  & 4.263 & 1.077 &  3.221 \\ 
             \hline
             10 & 4.263 & 1.077 &  3.221 \\
             15 & 4.263 & 1.078 &  3.219 \\
             20 & 4.263 & 1.078 &  3.218 \\
        \end{tabular}
        \end{ruledtabular}
    \end{table}
    \begin{figure}[t]
        \centering
        \includegraphics[width=0.48\textwidth]{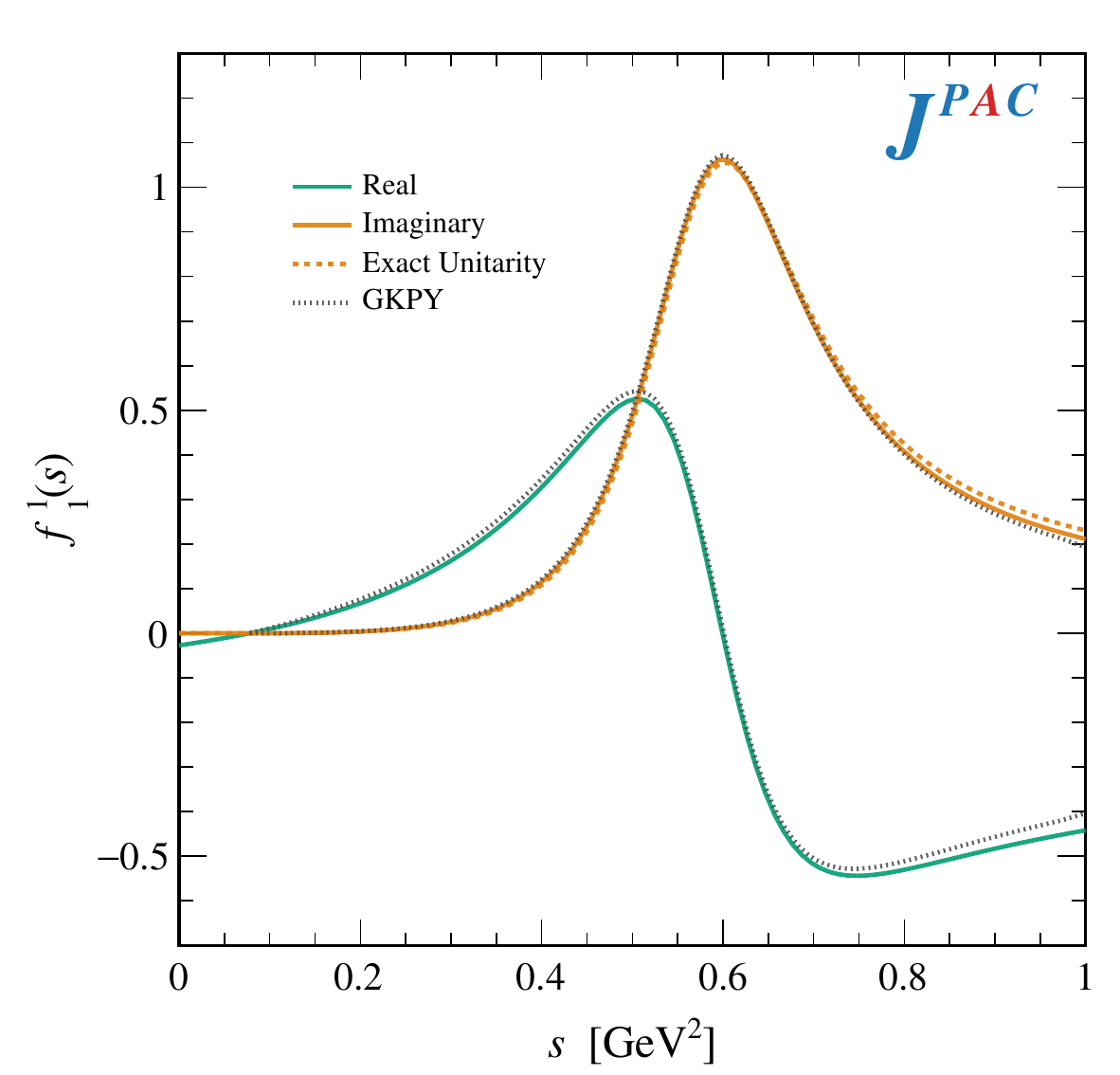}
        \caption{Best-fit of the $P$-wave projection of \cref{eq:F_1} in the elastic region (solid). For comparison, the GKPY PW (black dotted) is plotted and the expectation of elastic unitarity if it were satisfied exactly, i.e., \cref{eq:KT_homogeneous} (orange dashed).}
        \label{fig:f11}
    \end{figure}

\begin{figure}[t]
        \centering
        \includegraphics[width=0.48\textwidth]{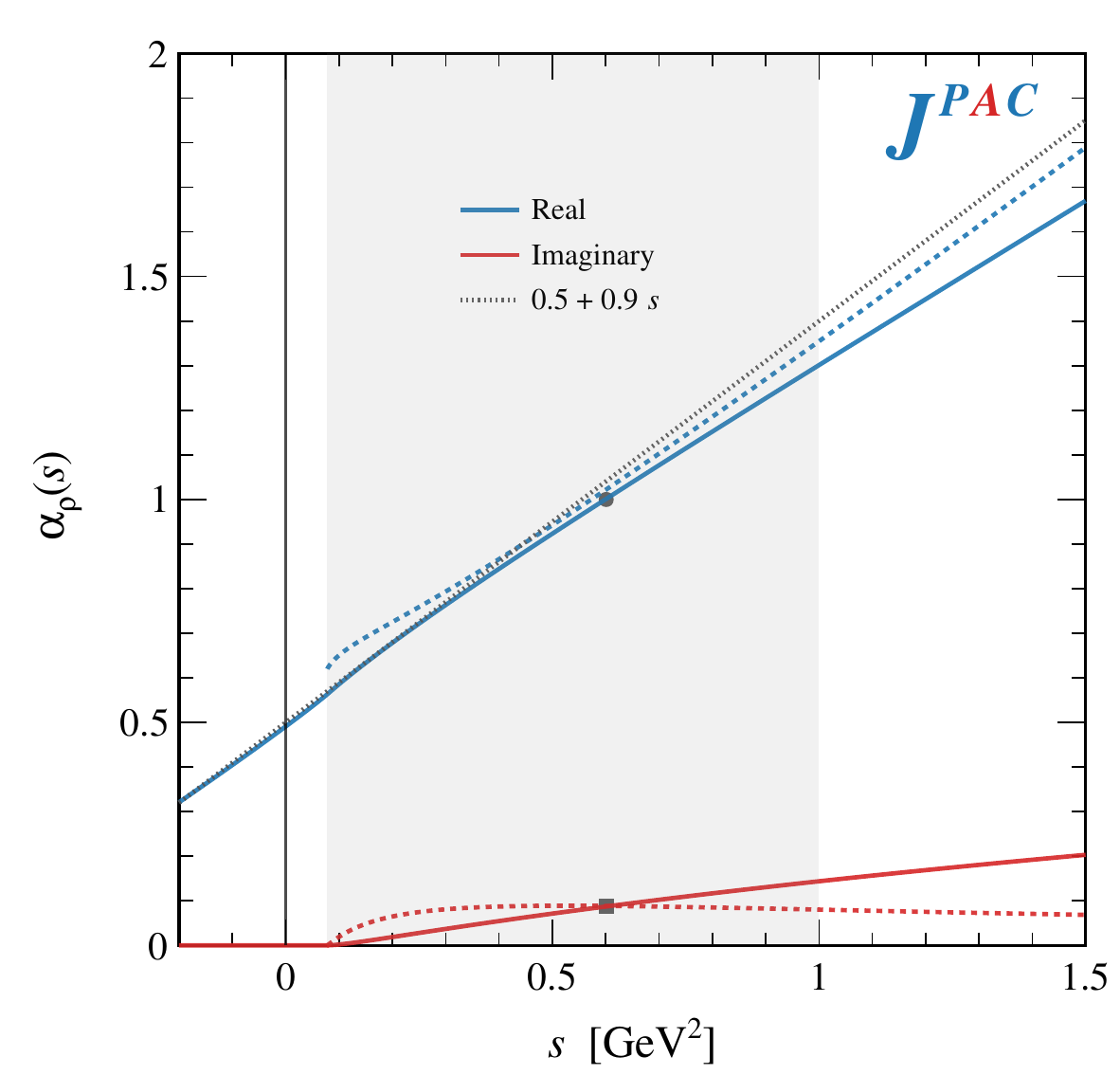}
        \caption{Best-fit trajectory $\alpha_\rho(s)$ after the iterative fitting procedure in the resonance region (solid) compared to the CRP trajectory from \cite{Londergan:2013dza} (dashed). The fitted region is shaded and the phenomenological trajectory $(0.5 + 0.9 \, s)$ (black dotted) is also plotted for comparison. The points correspond to $\Re \alpha(m_\rho^2) = 1$ (circle) and $\Im\alpha(m_\rho^2) = m_\rho \, \Gamma_\rho \, \Re\alpha^\prime(m_\rho^2)$ (square) at the nominal BW $\rho$ mass $m_\rho = 775\MeV$ and width $\Gamma_\rho = 148\MeV$.}
        \label{fig:rho_results_timelike}
    \end{figure}

The resulting $P$-wave is plotted in \cref{fig:f11} with parameters in \cref{tab:rho_pars}, where we see generally good agreement with unitarity and the GKPY amplitude. The resulting trajectory $\alpha_\rho(s)$ is plotted in the resonance region in \cref{fig:rho_results_timelike}, where $\alpha_\rho(s)$ crosses through the point $\Re\alpha(m_\rho^2) = 1$ as well as $\Im \alpha(m_\rho^2) = m_\rho \, \Gamma_\rho \, \Re\alpha^\prime(m_\rho^2)$, which is expected of the nearly BW nature of the $\rho$ lineshape. Note, however, that compared to the prototypical linear trajectory, the slope begins to decrease well within the fit region with deviations beginning just after the $\rho$ mass, i.e., $s \gtrsim 0.5\GeV^2$. We may also compare with other dispersive trajectories that incorporate unitarity, in particular the $\rho$ RT calculated in Ref.~\cite{Londergan:2013dza} using a ``constrained Regge pole" (CRP) model~\cite{Epstein:1968vaa,Chu:1968ctr}. We see a similar trend with the two coinciding near the $\rho$ mass.  We do note that the CRP trajectory has a larger imaginary part near threshold, in a region where the authors already observed that the PW amplitude is overestimated. This trajectory is calculated fixing only the complex pole position and not with a fit to the PW (i.e., the opposite approach to this analysis). Comparing the two methods, thus suggests that constraining the energy dependence of the residues in the numerator is important for extracting the RT in the denominator from fits to the PW amplitude.

    \begin{figure}[t]
        \centering
        \includegraphics[width=0.48\textwidth]{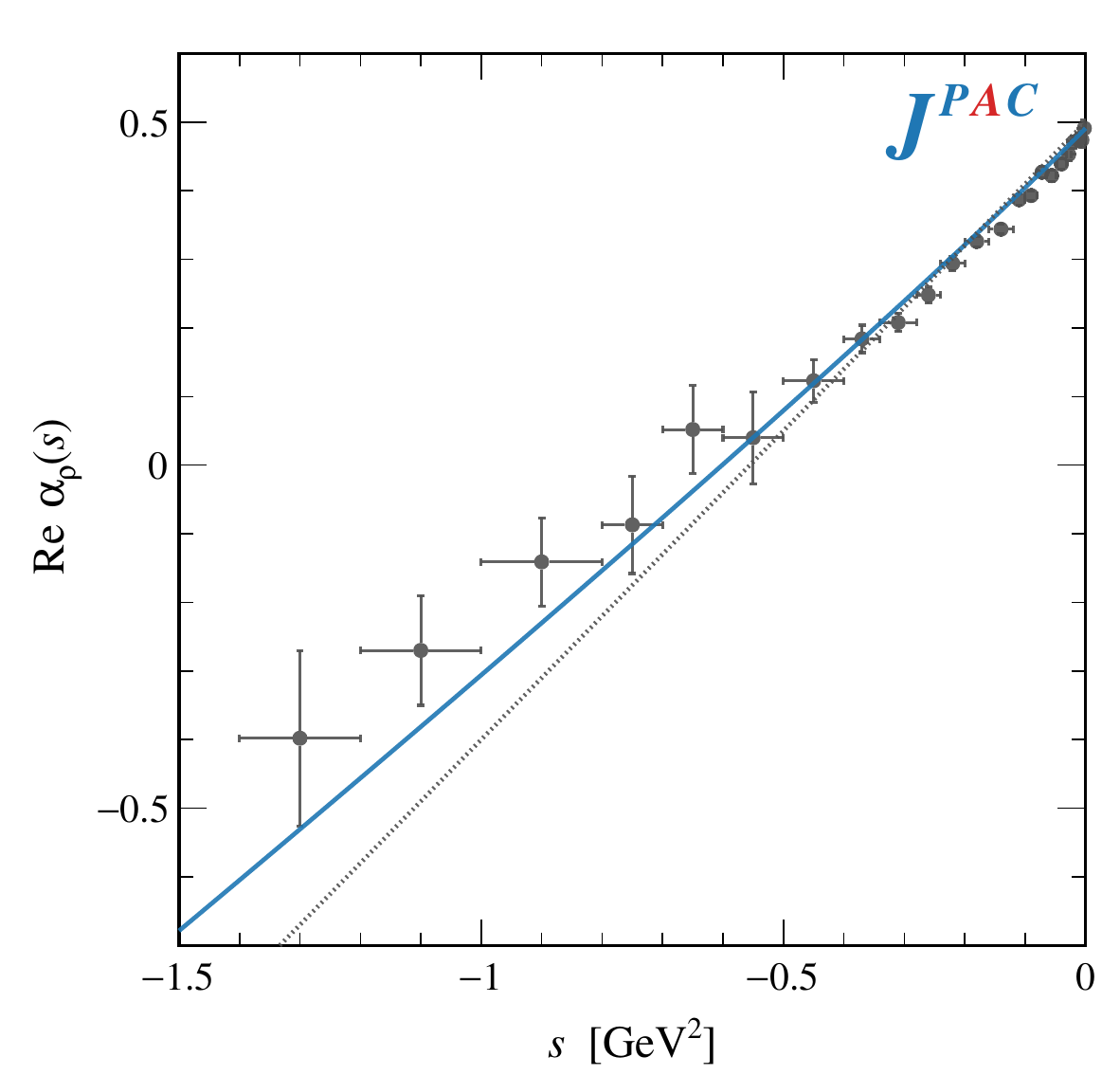}
        \caption{Best-fit trajectory $\alpha_\rho(s)$ plotted for spacelike energies. The curves are the same as \cref{fig:rho_results_timelike}. Data points are the effective $\rho$ trajectory as extracted from charge-exchange $\pi N$ scattering at high energies~\cite{Barnes:1976ek}. These data are shown for comparison but were not included in the fit.}
        \label{fig:rho_results_spacelike}
    \end{figure}

Because our RT is analytic, it can also be evaluated for spacelike energies below threshold, which is shown in \cref{fig:rho_results_spacelike}. We compare $\alpha_\rho(s)$ with the experimental extraction of the effective RT in charge-exchange $\pi N$ scattering~\cite{Barnes:1976ek}, which first observed the non-linearity of $\rho$ exchange at large momentum transfers. Remarkably, despite not being included in the fit, $\alpha_\rho(s)$, as constrained by elastic unitarity, is compatible with all data points. 

Finally, because the RT contains all the relevant information on the particle spectrum, $\alpha_\rho(s)$ can be used to examine the locations of resonance poles in the complex plane by searching for roots of $(j - \alpha(s))$ for \textit{any} positive integer $j$. Thus, in addition to the $\rho(770)$, we can extrapolate outside the fit range to extract the pole positions of the first radial excitation, $\rho(1450)$ or $\rho^\prime$, and first orbital excitation $\rho_3$, which are located at $\alpha_\rho(s_2) = 2$ and $\alpha_\rho(s_3) = 3$, respectively.
In the case of the former, the minimal model considered in \cref{eq:F_1} does not contain an explicit pole for this state as the $j=2$ term gets canceled by the signature factor. Ultimately, resonance masses and widths will only depend on the RT on which they appear and we can use the RT constrained around the $\rho(770)$ to predict the location of higher states. The structure of daughter poles in \cref{fig:chew_frautschi} means that the pole located at $\sqrt{s_3}$ will be degenerate with the second radial excitation, the $\rho(1700)$ or $\rho^{\prime\prime}$, which we also compare against. Details of the pole extraction are provided in \cref{app:poles}.
    \begin{table}[t]
        \centering
        \setlength{\tabcolsep}{4.5pt}
        \caption{Comparison of extracted complex roots of $\alpha_\rho(s_j) = j$ with the masses and widths of the observed $\rho$ spectrum. Only the RPP values for the $\rho(770)$ correspond to $T$-matrix pole parameters. The remaining quoted values are BW masses and widths.}
        \label{tab:rho_resonances}
        \begin{tabular}{c|cc|cc}
        \hline\hline 
             &  \multicolumn{2}{c|}{This work} & \multicolumn{2}{c}{RPP~\cite{ParticleDataGroup:2022pth}} \\
            \cline{2-5}
             & $\alpha_\rho(s_j)$ & $\sqrt{s_j}$ [MeV] &   $m_j$ [MeV] & $\Gamma_j/2$ [MeV] \\
             \hline
             $\rho(770)$ & 1 & $760 - i\, 70$ & (761 -- 765) & (71 -- 74) \\
             \hline
             $\rho(1450)$ & 2 & $1380 - i 120$ & $1465\pm 25$ & $200\pm 30$ \\
             \hline
             $\rho_3(1690)$ & \multirow{2}{*}{3} & \multirow{2}{*}{$1800 - i 130$} & $1688\pm 2.1$ & $80\pm 5$  \\
             $\rho(1700)$ & &&  $1720 \pm 20$ & $125\pm 50$ \\
             \hline\hline
         \end{tabular}
    \end{table}

The resulting pole positions are tabulated in \cref{tab:rho_resonances}, where we see the parameters of the $\rho(770)$ are generally in good agreement with more detailed dispersive analyses~\cite{Garcia-Martin:2011nna,Hoferichter:2023mgy,Heuser:2024biq}. In addition, using \cref{eq:im_pw_rho,eq:residue} we extract the modulus and phase of the residue
    \begin{equation}
        |g_{\rho\pi\pi}| = 5.8 \quad \text{ and } \quad \phi_{\rho\pi\pi} = -6.2\degree  ~,
    \end{equation}
which are also in qualitatively good agreement~\cite{Garcia-Martin:2011nna,Hoferichter:2023mgy,Heuser:2024biq}. We present these values as well as the pole positions without error analysis as we aim only for an exploratory benchmark of the isobar and trajectory models. 

The higher poles of the $\rho^\prime$, $\rho^{\prime\prime}$, and $\rho_3$ also compare reasonably well with the RPP masses and widths in \cref{tab:rho_resonances}, but deviate more than the ground state $\rho$. This could be due to several factors: first, the quoted RPP values for these states correspond to BW masses and widths. Since these states are generally broader and harder to extract than the $\rho$, there may be substantial deviations from the genuine pole location. Second, the model in \cref{eq:mF_resonance} is not unitary by construction and thus extrapolating far outside the fit range may suffer violations from unitarity. Finally, as previously mentioned, higher $\rho$ resonances couple primarily to inelastic channels such as $4\pi$. Since the RT couples to multi-body channels in a possibly non-trivial way, our trajectory may be ignoring important effects from inelastic thresholds.

We do not attempt to quantify the uncertainties from these effects here. Nevertheless, finding resonance poles located in generally the right place in the complex plane, even when extrapolated far from the fit region, is reassuring and a first step to a more in-depth exploration of these poles in the future.  

%%%%%%%%%%%%%%%%%%%%%%%%%%%%%%%%%
\subsection{\texorpdfstring{$I=0$}{I=0} and \texorpdfstring{$\alpha_\sigma(s)$}{alpha sigma(s)}}
\label{sec:I=0}
%%%%%%%%%%%%%%%%%%%%%%%%%%%%%%%%%

Turning to the $I=0$ amplitude, we wish to investigate the $\sigma/f_0(500)$ resonance, which is seen in the $S$-wave near-threshold alongside the narrow $f_0(980)$. In our formalism, this channel will receive contributions from all RTs of similar quantum numbers and one must also account for the $f_2$ and $\mathbb{P}$ trajectories. We therefore write 
\begin{widetext}
    \begin{align}
        \label{eq:F_0}
        \mF{0}(s,z_s) &= \frac{1}{2} \bigg[ g_{\sigma}^2\, F(\alpha_{\sigma}(s), \, \nu_s) + g_{f_2}^2 \, F(\alpha_{f_2}(s), \, \nu_s) + g_\mathbb{P}^2 \, F_\text{NR}(\aspom,  \, \nu_s) \bigg] + (\nu_s \leftrightarrow - \nu_s) ~,
    \end{align}
\end{widetext}
where the $\sigma$ and $f_2$ use the resonant isobar in \cref{eq:f(alpha_z)} and the Pomeron is non-resonant using \cref{eq:f(alpha_z)_Pomeron}.
Only a single trajectory for the scalar resonances is included as we will restrict our fitting to energies close to threshold in order to focus on the $\sigma$. Follow-up investigations about the $f_0(980)$ can be conducted by including an additional term in \cref{eq:F_0}.

We first discuss the $f_2$ trajectory, which is expected to yield the largest contribution resulting from $I=0$ hadron exchanges at high energies, subleading only to the exchange of a Pomeron. The smallness of the $\pi\pi$ scattering cross sections at maximal isospin is often explained by cancellations between the Reggeons in the crossed channel due to the (approximate) exchange degeneracy of the $\rho$ and $f_2$ trajectories~\cite{Lipkin:1969dh}. Specifically, if 
    \begin{equation}
        \label{eq:EXD}
        \alpha_\rho(s) \approx \alpha_{f_2}(s) \quad \text{ and } \quad g^2_{f_2} \approx \frac{3}{2} \, g^2_\rho ~,
    \end{equation}
then the sum of imaginary parts from the $\rho$ and $f_2$ in \cref{eq:mA_I_regge} will vanish asymptotically. We may do a simple test of exchange degeneracy using the $\alpha_\rho(s)$ as calculated in the previous section by comparing the quoted $T$-matrix poles of the $f_2(1270)$ and its degenerate daughter pole $f_0(1370)$~\cite{ParticleDataGroup:2022pth} 
    \begin{subequations}
    \begin{align}
        \sqrt{s_{f_2(1270)}} &= (1260 \text{ -- } 1283) - i \, (90 \text{ -- } 110) ~,  \\
        \sqrt{s_{f_0(1370)}} &= (1250 \text{ -- } 1440) - i \, (60 \text{ -- } 300) ~.
    \end{align}
    \end{subequations}
\Cref{eq:EXD} suggests these should be compared with the $\alpha_\rho(s_2) = 2$ entry in \cref{tab:rho_resonances}, which once again compare rather well, but without a detailed error analysis no definite conclusions can be drawn.
In any case, because the first physical resonance on the $f_2$ trajectory has spin-2, this isobar must have $\jmin = 1$ or $2$ in order to avoid the pole at $j=0$.
At low energies the contributions of the $f_2$ Regge pole will always be $\O{\qhats^4}$ as dictated by \cref{eq:mF_resonance}, and details of the trajectory and coupling are largely suppressed when considering very near threshold energies.
Thus, taking the $S$-wave PW projection of \cref{eq:F_0} and considering only leading order in the momentum expansion yields
    \begin{equation}
        \label{eq:f00_before_simplify}
        f_0^0(s) = \frac{g_{\sigma}^2}{-\alpha_{\sigma}(s)} - \frac{g_\mathbb{P}^2}{\Gamma(2-\alpha_\mathbb{P}(s))} + \O{\qhats^4} ~,
    \end{equation}
without a contribution from the $f_2$. Because we are ignoring the crossed channels in the homogeneous unitarity equation, the Pomeron coupling and trajectory are largely unconstrained. For simplicity then, given the limited energy range considered, we will assume $\alpha_\mathbb{P}(s) \approx \alpha_\pom(0) \simeq 1$ and fit the coupling $g_\mathbb{P}^2$. Since our primary goal is the extraction of $\alpha_\sigma(s)$ we keep the full tower of poles in the hypergeometric function for $F(\alpha_\sigma(s), \nu_s)$ and absorb all other $\O{\qhats^4}$ and higher contributions of both the $\pom$ and $f_2$ in \cref{eq:F_0} into the fitted coupling $g_\pom^2$. In a fully crossing symmetric analysis, the value of the coupling $g_\pom^2$ should instead be compared with the total $\pi^+\pi^-$ cross section at high energies using \cref{eq:mA_I_regge}.

With these simplifications in mind, we approximate \cref{eq:F_0} as 
\begin{subequations}
    \begin{align}
        \label{eq:F_0_fit}
        \mF{0}(s,z_s) &= \frac{1}{2} \bigg[ g_{\sigma}^2\, F(\alpha_{\sigma}(s), \, \nu_s) - g_\mathbb{P}^2\bigg] + (\nu_s \leftrightarrow - \nu_s) ~,
    \end{align}
such that the resulting $S$-wave projection is written as
    \begin{equation}
        \label{eq:f00}
        f_0^0(s) = \frac{g_{\sigma}^2 + g_\mathbb{P}^2 \, \, \alpha_{\sigma}(s)}{-\alpha_{\sigma}(s)} + \O{\qhats^4} ~.
    \end{equation}
\end{subequations}
Note that because $g_{\sigma}^2$ and $g_\mathbb{P}^2$ are both real and positive, the numerator will manifest a zero if for some real \mbox{$s_A < 4 \, m_\pi^2$}, the trajectory satisfies $\alpha_{\sigma}(s_A) = - (g_{\sigma}/g_\mathbb{P})^2$. This is the Adler zero, required by chiral symmetry~\cite{Adler:1964um,Adler:1965ga}. We remark that the existence of such a zero in the chiral limit, i.e., at $s_A = 0$, requires $\alpha_{\sigma}(0) \leq 0$ and would be consistent with a quickly vanishing Regge exchange contribution at high energies, but we make no \textit{a priori} assumptions about the location of the zero. 

We briefly comment on how the Adler zero arises in \cref{eq:f00} as compared to the Veneziano--Lovelace--Shapiro (VLS) model~\cite{Veneziano:1968yb,Lovelace:1968kjy,Shapiro:1969km}. In the latter, requiring a zero implies a relation between the trajectories appearing in different channels. Since only a single trajectory was considered (i.e., that of the exchange degenerate $\rho-f_2$ mesons), this required $\alpha(s) + \alpha(t) = 1$ and fixes $\alpha(0) = 1/2$ in the chiral limit. From this perspective, the zero is a manifestly crossing symmetric phenomenon with the direct and crossed channel RTs interfering near the Adler point. In \cref{eq:f00}, the zero arises from the interplay of purely $I=0$ trajectories in the direct channel and necessitates a trajectory with negative intercept, a typical feature of scalar resonances, interfering with the Pomeron. Considering subleading terms in the momentum expansion, more trajectories, or the fully crossing symmetric combination \cref{eq:mA_I} can modify the location of the zero, but will not change the basic mechanism of \cref{eq:f00}. While the original VLS model cannot accommodate the Pomeron, extensions to include its effects concluded it should play an important role in satisfying chiral constraints~\cite{Huang:1969ef,Wong:1969bt}. As such, we find the interpretation of the zero in terms of $\sigma/\pom$ interference in the direct channel particularly appealing.\footnote{The $I=2$, $S$-wave is also predicted to have an Adler zero as a consequence of chiral symmetry. In our formalism, this zero must come from a different mechanism than that of $I=0$ as, in the former, the Pomeron only contributes indirectly through the crossed channel. We do not investigate plausible alternatives for this channel here.}

Further, since unitarity will only affect the shape of the RT, the amplitude \cref{eq:f00} will contain an Adler zero even if we only consider the homogeneous unitarity equation without requiring the inverse amplitude, in this case $\alpha_\sigma(s)$, to have a sub-threshold pole. This is unlike the typical Omn\`es function approach, which requires introducing \textit{ad hoc} parameters, i.e., subtraction polynomials~\cite{Danilkin:2022cnj}, when considering homogeneous unitarity.

We may enforce \cref{eq:KT_homogeneous} at leading order of \cref{eq:f00} if 
    \begin{align}
        \label{eq:f0_unitarity}
        \Im \alpha_{\sigma}(s)=  
        \rho(s) \left[ g_{\sigma}^2 \left|  1 + \frac{g^2_\mathbb{P}}{g^2_\sigma}
        \alpha_{\sigma}(s) \right|^2 + \O{\qhats^4} \right] ~.
    \end{align}
Since by assumption $|\alpha_{\sigma}(s)| < s$, \cref{eq:f0_unitarity} is still of the form in \cref{eq:log_sth,eq:rs}, albeit no longer a simple polynomial in momentum.

We begin by assuming that the $\sigma$ RT admits a solution similar to the $\rho$ as considered in \cref{sec:I=1}. If this is the case, $\Re\alpha_\sigma \sim \sqrt{s} \to + \infty$ with known asymptotic limits for both real and imaginary parts and one may choose 
    \begin{align}   
        \label{eq:r_sigma}
        r_\sigma(s) &= 
        g_\sigma^2  \left|  1 + \frac{g^2_\mathbb{P}}{g^2_\sigma}
        \, \alpha_\sigma(s) \right|^2 + c_\sigma \, \qhats^{2(1+\Re\alpha_\sigma(s))} ~,
    \end{align}
such that 
    \begin{equation}
        \label{eq:f00_unitarity}
        \Im f_0^0(s) = \rho(s) \, |f_0^0(s)|^2 + \O{\qhats^{2(1+\Re\alpha_\sigma(s))}} ~,
    \end{equation}
satisfies \cref{eq:KT_homogeneous} at leading order assuming $\Re\alpha_\sigma(s \gtrsim 4m_\pi^2)$ is not too negative. 
Since very little is known about the $\sigma$ trajectory from measurements, we have more parameters, i.e., $\alpha_\sigma(0)$, $g^2_\sigma$, $g^2_\pom$, $\gamma_\sigma$, and $c_\sigma$, to determine by minimizing \cref{eq:leastsquares} with our iterative fitting procedure. We choose an initial guess of $\alpha_\sigma(s) = (-0.2 + 0.1 \, s)/\sqrt{1+s/20}$ and fit evenly spaced points from $4m_\pi^2$ to $0.5\GeV^{2}$. This fitting range is selected to minimize the effect of the $f_0(980)$, which is not included. 

\begin{figure}[t]
        \centering
        \includegraphics[width=0.48\textwidth]{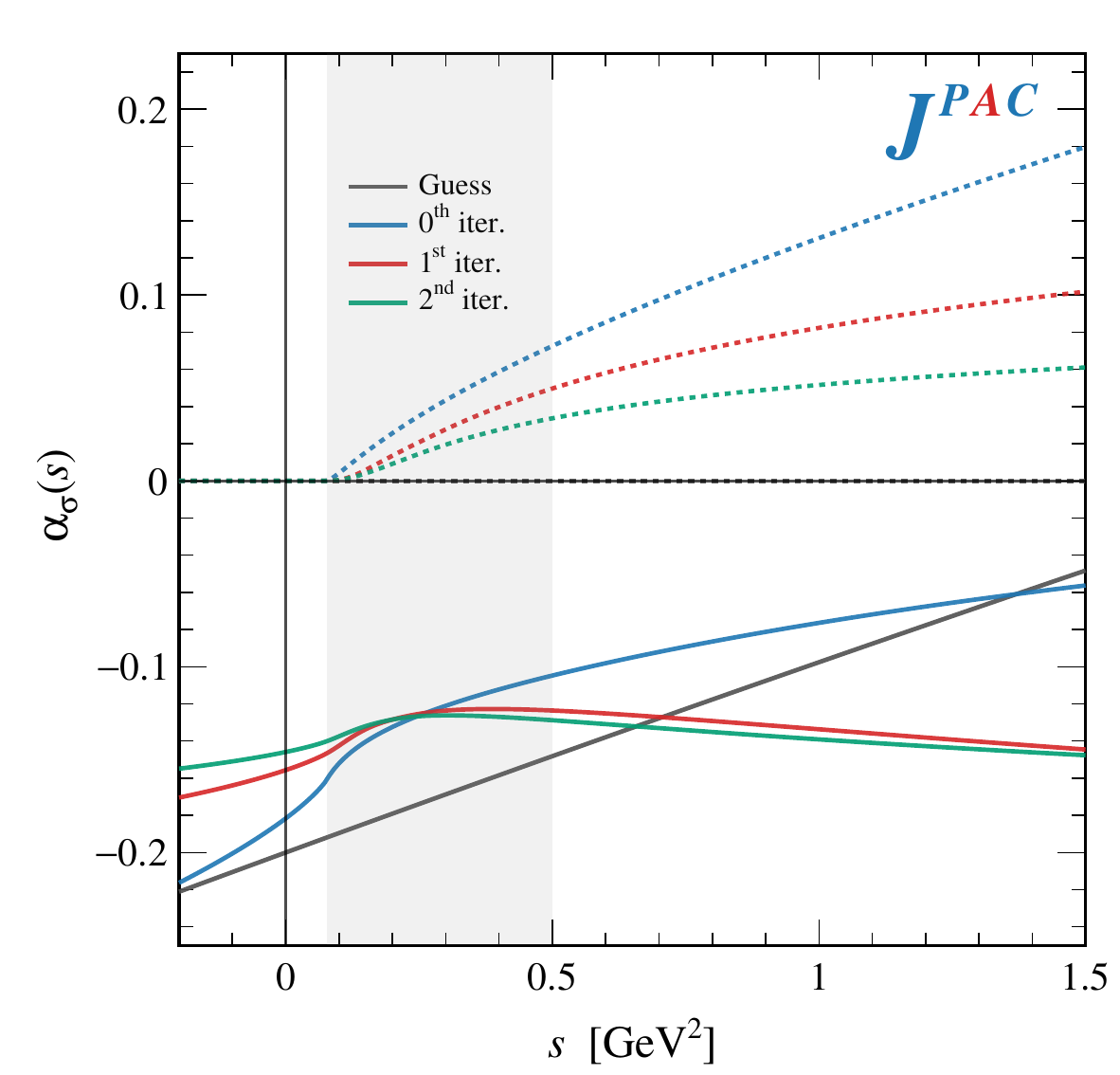}
        \caption{First three iterations of the integral equation for $\Re\alpha_\sigma(s)$ (solid) and $\Im\alpha_\sigma(s)$ (dashed) using \cref{eq:r_sigma}. The fitted region is shaded and we see a trend towards negative $\Re\alpha_\sigma(s)$  with subsequent iterations.}
        \label{fig:alpha_sigma_iters}
    \end{figure}

The first two iterations are plotted in \cref{fig:alpha_sigma_iters}, where we see a dramatically different behavior than in \cref{fig:rho_results_timelike}. Specifically, subsequent iterations have the effect of reducing the slope of $\Re\alpha_\sigma(s)$ as it tends to negative values. At the same time, the parameters at each iteration tabulated in \cref{tab:sigma_pars} reveal $c_\sigma$ approaches zero and, thus, the term $\propto \qhats^{2\Re\alpha_\sigma(s)}$, which is responsible for the indefinite rise of the real part of the RT, decouples from the integral equation. This implies that elastic unitarity disfavors stable solutions for the $\sigma$ of the same type as for the $\rho$, with the former preferring an imaginary part which grows like some power of $\log s$ and $\Re\alpha_\sigma(s) \to - \infty$. Different choices of the initial guess change the other parameters and rate of $c_\sigma \to 0$, but ultimately reach the same conclusion.
    \begin{table}[t]
        \caption{Fit parameters for the $\sigma/f_0(500)$ trajectory. We tabulate the first two iterations resulting from using \cref{eq:r_sigma} as well as the non-iterative solution using \cref{eq:rhat_sigma}.}
        \label{tab:sigma_pars}
        \centering
        \begin{tabular}{ccccccc}
             \hline \hline 
             $i$ & $\alpha(0)$ & $g_\sigma^2$ & $g^2_\pom$ & $\gamma_\sigma$ & $c_\sigma$ & $c_\sigma^\prime$\\
             \hline 
             \multicolumn{7}{c}{[\cref{eq:r_sigma}]} \\
             \hline 
             0 & $-0.182$ & $0.228$ & $1.170$ & $0.523$ & $1.053$ &  \\            
             1 & $-0.156$ & $0.362$ & $2.243$ & $0.111$ & $3.78 \cdot 10^{-3}$ & \\
             2 & $-0.146$ & $0.533$ & $3.586$ & $0.082$ & $2.11 \cdot 10^{-5}$ &  \\
             \hline 
             \multicolumn{7}{c}{[\cref{eq:rhat_sigma}]} \\
             \hline 
              &  & $0.137$ & $1.886$ & &  & $0.014$ \\
             \hline \hline 
        \end{tabular}
    \end{table}

Already, we can see that the shape being approached in \cref{fig:alpha_sigma_iters} is highly non-linear and resembles the RTs from potential theory in non-relativistic quantum mechanics rather than the quintessential ``stringy" dynamics of a relativistic confined quark model~\cite{Mandelstam:1969dk}. In addition, none of the iterations in \cref{fig:alpha_sigma_iters} cross zero near threshold implying that, if these contain the $\sigma$ pole, it is located deep in the complex plane and unlike a typical BW. 

With the previous considerations, it is possible to choose a different parameterization by setting $c_\sigma = 0$ in \cref{eq:r_sigma} from the beginning. Since we anticipate $\Re\alpha(s) \to -\infty$, \cref{eq:Imalpha_bound} will be trivially satisfied and it is no longer required to find a stable iterative solution with $\gamma_\sigma >1$. 
In fact, no iterative solution is needed at all, by choosing 
    \begin{align}
        \label{eq:rhat_sigma}
        \Im\alpha_\sigma(s)& = \rho(s) \, g^2_\sigma \, 
        \\
        &\hspace{-1cm}\times \left(1 + \frac{g^2_\mathbb{P}}{g^2_\sigma}\left[\alpha_\sigma(4m_\pi^2) + c^\prime_\sigma \, \rho(s) \, \log\left(\frac{s}{4m_\pi^2}\right)
        \right]\right)^2
        \nonumber ~, 
    \end{align}
which is derived from \cref{eq:r_sigma,eq:log_form} by taking \mbox{$\gamma_\sigma \to \infty$}. This limit is taken because, with $c_\sigma =0$, the trajectory will be asymptotically logarithmic (cf.\ \cref{eq:s_ininity_log}) and having $\Im\alpha_\sigma(s) \propto |\alpha_\sigma(s)|^2$ will no longer spoil the convergence of \cref{eq:disp_one_sub}. We may thus remove the overall logarithm of \cref{eq:log_form} and dependence on $\gamma_\sigma$ with the aforementioned limit. By shifting the subtraction point of the dispersion integral to $s=4m_\pi^2$, we may further approximate the factor of $\alpha_\sigma(s)$ on the right-hand side of \cref{eq:f0_unitarity} by $\alpha_\sigma(s) \simeq \alpha_\sigma(4m_\pi^2) + c_\sigma^\prime \, \rho(s) \, \log (s/4m_\pi^2)$,\footnote{Note that the  \cref{eq:rhat_sigma} vanishes at threshold as $\Im\alpha_\sigma(s) \propto \rho(s) \propto \sqrt{s-4m_\pi^2}$ and, thus,  $\Re \alpha_\sigma(s) \to \alpha_\sigma(4m_\pi^2)$ linearly in $(s-4m_\pi^2)$. The empirical simplification converges to the threshold value faster, but this does not affect the analytic properties of $\alpha_\sigma(s)$.} which empirically describes the near-threshold behavior. 
Strictly speaking, since $\alpha_\sigma(s)$ is complex-valued above threshold, the parameter $c^\prime_\sigma$ should also be complex. The imaginary part must be a real function, however, and we keep $c_\sigma^\prime$ as a real parameter in \cref{eq:rhat_sigma}, to be determined by the fit, and drop the absolute value. 

Using \cref{eq:rhat_sigma}, in principle, four free parameters are left to be determined by fit: $g_\mathbb{P}^2$, $g_\sigma^2$, $\alpha_\sigma(4 m_\pi^2)$, and $c_\sigma^\prime$. Note, however, that because the RT is now given in terms of fixed parameters instead of an iterative interpolation, up to an overall constant, the PW in \cref{eq:f00} will only depend on the ratios $\alpha_\sigma(4m_\pi^2)/g_\sigma^2$ and $c^\prime_\sigma/g_\sigma^2$. Then, since these are all \textit{a priori} undetermined, at least one will be redundant and we fix $\alpha(4 m_\pi^2) = -0.064$. This is the threshold value of the $\sigma$ RT as calculated using the CRP model in Ref.~\cite{Londergan:2013dza}, allowing for a direct comparison between the two approaches. When considering the fully crossing symmetric model, which is left for a future extension of this work, the parameters of the isobars in the crossed channel, i.e., the $\rho$ coupling, will set a scale to break this ambiguity and allow a determination of $\alpha_\sigma(4 m_\pi^2)$ (or equivalently $\alpha_\sigma(0)$).

\begin{figure}[t]
        \centering
        \includegraphics[width=\linewidth]{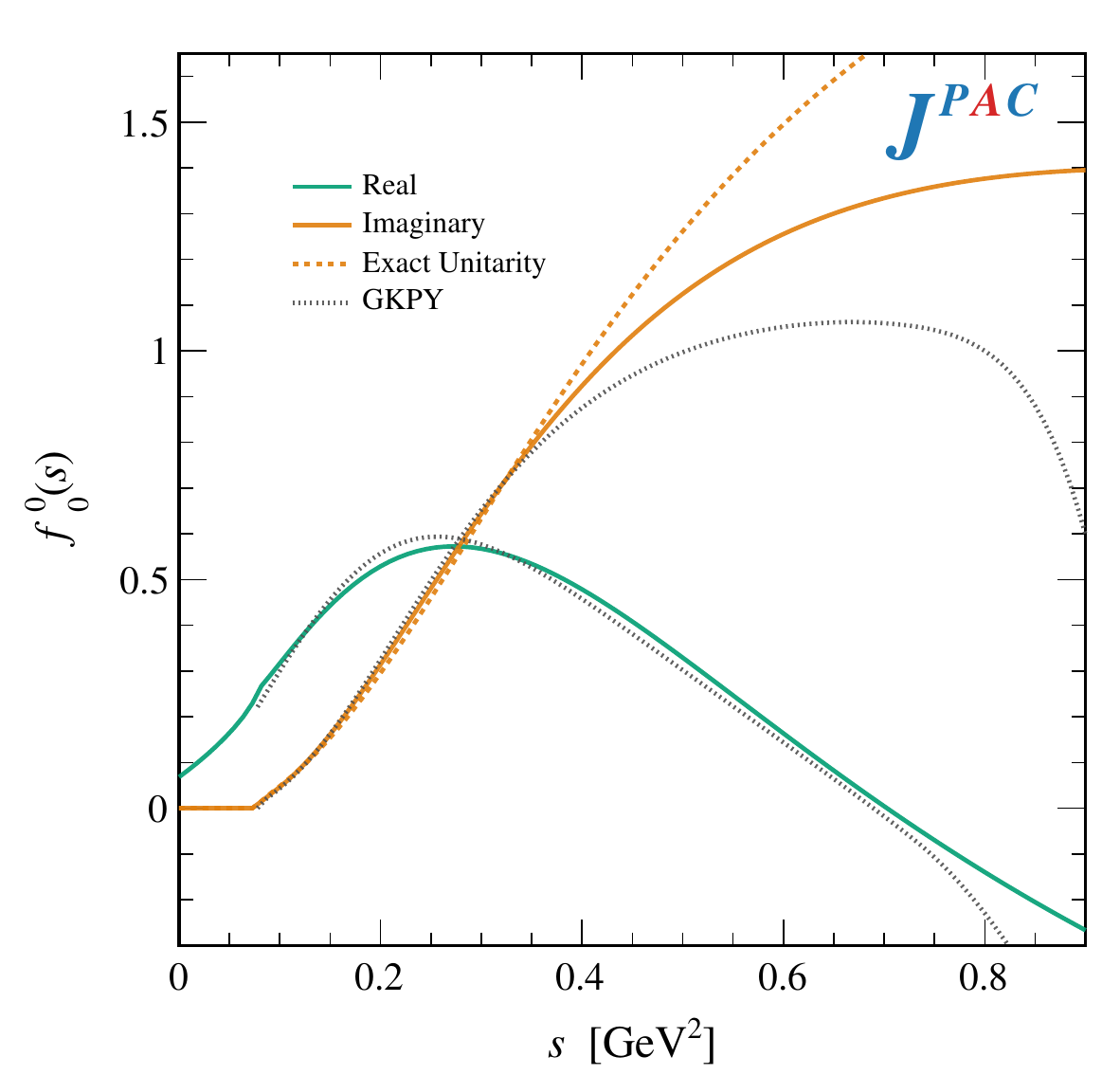}
        \caption{The $S$-wave PW projection for the $I=0$ amplitude using \cref{eq:f00_before_simplify} as described in the text. Legend is the same as in \cref{fig:f11}.}
        \label{fig:f00_results}
    \end{figure}

The resulting $S$-wave amplitude and its parameters are shown in \cref{fig:f00_results} and \cref{tab:sigma_pars}, respectively. Generally good agreement with unitarity within the fit range is observed, but significant deviations appear when extrapolating outside. As discussed before, higher-order terms in momentum, the presence of the $f_0(980)$, and the $K\bar{K}$ threshold become important at $s \geq 0.5\GeV^2$ and are beyond the scope of our minimal model.

\begin{figure}[t]
        \centering
        \includegraphics[width=\linewidth]{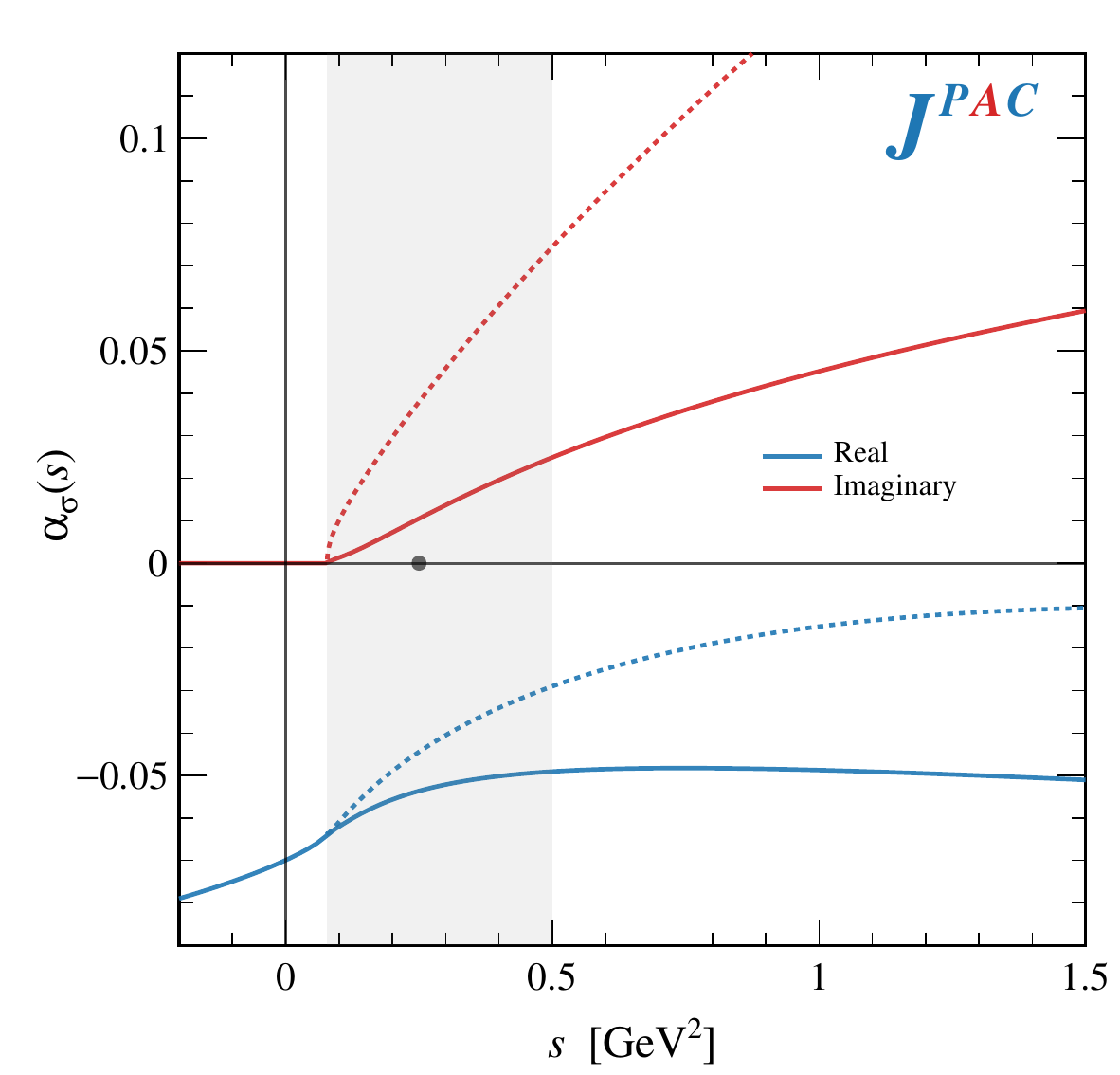}
        \caption{The $\sigma$ trajectory as constrained by homogeneous elastic unitarity in \cref{eq:KT_homogeneous} using \cref{eq:rhat_sigma} (solid). The fitted region is shaded and we compare with $\alpha_\sigma(s)$ as calculated in Ref.~\cite{Londergan:2013dza} (dashed) as well as with the expectation of a BW-like $\sigma$ satisfying $\Re\alpha_\sigma(m_\sigma^2 \sim (0.5 \GeV)^2) = 0$ (grey dot). Unlike \cref{fig:rho_results_timelike}, no corresponding point calculating $\Im\alpha_\sigma(s)$ via \cref{eq:BW_width} is shown.}
        \label{fig:sigma_results}
    \end{figure}

Examining the RT in \cref{fig:sigma_results}, we see a very similar shape to that being approached in \cref{fig:alpha_sigma_iters}: a nearly flat real part, which eventually turns downward before diverging to $\Re\alpha(s) \to -\infty$ logarithmically. We compare explicitly with the real and imaginary parts of the CRP $\alpha_\sigma(s)$ in Ref.~\cite{Londergan:2013dza}. Therein, a twice-subtracted dispersion relation, the $S$-wave PW with a single pole, and elastic unitarity are used and are in qualitatively good agreement with the GKPY amplitude. The CRP analysis concluded that the $\sigma$ pole was consistent with a RT with a very small slope. In comparison, our trajectory in \cref{fig:sigma_results}, grows even slower and is thus also consistent with a $\sigma$ meson featuring a non-ordinary Regge behavior. Another indication of this is the observation that the real part of neither RT in \cref{fig:sigma_results} crosses the real axis in the vicinity of the $\sigma$ meson. As was seen with the $\rho$ resonance, the mass and width of narrow resonances lying on conventional rising trajectories can be well estimated by expanding the RT around the integer values of the real part. This is clearly not the case for the $\sigma$ whose real part has no zero crossing. 

While our free parameters are determined by fitting the PWs, the CRP trajectory was calculated by ensuring that the $\sigma$ pole and residue in the complex plane are consistent with those quoted in Ref.~\cite{Garcia-Martin:2011iqs}. In order to compare, we also continue our trajectory to the complex plane where a single root satisfying $\alpha(s_\sigma) = 0$ is found with
    \begin{equation}
        \sqrt{s_\sigma} = (500 - i \, 470)\MeV ~,
    \end{equation}
and a residue in the PW of
    \begin{equation}
        |g_{\sigma\pi\pi}| = 7.9\GeV 
        \quad \text{and} \quad
        \phi_{\sigma\pi\pi} = -72\degree ~.
    \end{equation}
While the mass and phase are in qualitatively good agreement with typical results extracted from precision studies~\cite{Garcia-Martin:2011nna,Hoferichter:2023mgy,Danilkin:2020pak,Albaladejo:2012te,Rodas:2023nec}, the width and size of the coupling deviate by about a factor of two.
This is not entirely unexpected, as we recall that the model in \cref{eq:f00}, which is used to constrain the RT, is devoid of $s$-dependence except that coming from the $\alpha_\sigma(s)$. More specifically, using the homogeneous unitarity equation \cref{eq:KT_homogeneous}, we ignore any contribution from LHCs, which are well known to be important contributions to $\sigma$ pole determinations~\cite{Xiao:2000kx,Caprini:2008fc,Pelaez:2021dak,Danilkin:2022cnj}. 
Because of this, $\alpha_\sigma(s)$ and therefore the width of the $\sigma$, must encompass the entire $s$-dependence of the PW, leading to this overestimation. 

Another indication of this effect is the location of the Adler zero, which is located at 
$s_A = -0.05\GeV^2$. From ChPT predictions, we should expect the Adler zero to be some \textit{positive} factor of $m_\pi^2$ below threshold~\cite{Danilkin:2022cnj}. While the leading-order prediction places the zero at $s_A = m_\pi^2/2$~\cite{Weinberg:1966kf}, its precise location is typically sensitive to the implementation of the LHC given the close proximity of the branch point at $s=0$ as well as that of the RHC at $s = 4m_\pi^2$~\cite{Danilkin:2022cnj,Heuser:2024biq}. Our Adler zero at $s < 0$ would, in principle, overlap with the LHC if it were included and we thus expect a more in-depth analysis, which includes the full crossed channel contributions, to shift both pole and zero to values more consistent with other approaches. 
Regardless, we consider finding a pole (and zero) in the correct mass range with such a simple model and with few parameters provides reassurance that the formalism presented here is capable of describing more than just the classical examples of ``simple" resonances such as the $\rho$. 

One obvious advantage of working with RTs is that one can examine other properties about the $\sigma$, which would not be possible with typical PW analyses. Specifically, just like the extraction of the $\rho^\prime$ and $\rho_3$ in \cref{sec:I=1}, the existence of radial and orbital excitations of the $\sigma$ can be examined. These would-be $\sigma^\prime$ and $\sigma_2$ mesons correspond to the roots $\alpha_\sigma(s_{\sigma^\prime}) -1 = 0$ and $\alpha_\sigma(s_{\sigma_2})-2=0$, respectively, and can be searched in the same way as the ground state $\sigma$. Calculating $\alpha_\sigma(s)$ in the complex plane, we notice that on the unphysical sheet $\Re\alpha_\sigma(s) \to -\infty$ monotonically (and logarithmically) in every direction in the lower-half complex plane. Because of this, there are no positive integer values of $\alpha_\sigma(s)$ in the complex plane and we can conclude that the $\sigma$ does not have any excitations. This is once again more akin to the RT of a non-relativistic potential with finitely many bound states as opposed to a conventional Reggeon~\cite{Londergan:2013dza,Mandelstam:1969dk}. While it may be possible that coupling to higher thresholds can produce a RT that turns around above some energy, such a scenario would entail the excitations are somehow ``induced" by inelastic thresholds, which we find highly unlikely. 

%%%%%%%%%%%%%%%%%%%%%%%%%%%%%%%%%
\section{Summary and Outlook}
\label{sec:conclusions}
%%%%%%%%%%%%%%%%%%%%%%%%%%%%%%%%%
To summarize, in this work we presented a parameterization for scattering amplitudes that recovers many expectations of $S$-matrix principles both at low energies and in the asymptotic limit. Since analyticity and crossing necessarily correlate the physics of hadron interactions at all energies and in all partial waves, our proposed formalism in terms of hypergeometric isobars and Regge trajectories allows these constraints to be rigorously imposed by using relatively simple functional forms. In this way, our model is a realization of the unification of a wide array of QCD phenomena: resonances in the complex energy plane, existence of orbital/radial excitations, Reggeon exchanges in peripheral reactions, and even the scaling behavior emerging from parton exchanges in perturbative regimes.
We focused here on the scattering of identical, spinless particles, but this constitutes a starting point for further applications, generalizations, and improvements on the basic formalism. 

As first-order validation of the feasibility of our model, we examined low-energy $\pi\pi$ scattering. We showed the flexibility of our formalism to generalize for different isospins and meson families. A systematic way to enforce unitarity up to some momentum scale of interest, which closely parallels the implementation of the KT equations, was proposed and the Regge trajectories of the $\rho$ and $\sigma$ mesons were examined. Our results compared reasonably to existing literature and highlighted future improvements, which can lead to better extractions of these Regge trajectories. Specifically, we have paved the way towards generalizing the unitarization scheme to include the full effect of left-hand cuts in a crossing symmetric KT-like or Roy-like analysis using Regge poles. Because our formalism aims to describe scattering at all energies, such an analysis can also include experimental high-energy data directly to further constrain the amplitude. 

Further generalizations to decay processes, coupled channels, unequal masses, and particles with spin will greatly increase the applicability of our model and allow a much more constrained determination of Regge trajectories.  Since these are generalizations of poles in the relativistic $S$-matrix, Regge trajectories can be an important tool for the study of hadronic resonances.

%%%%%%%%%%%%%%%%%%%%%%%%%%%%%%%%%
\acknowledgments
We thank \mbox{Bastian~Kubis} and \mbox{Jos\'{e}~R.~Pel\'aez} for enlightening discussions and useful comments on the manuscript. We also thank \mbox{Miguel~Albaladejo} for instructive discussions in the early stages of this project. This work was initiated within the ``Scattering Theory" course as part of the Indiana University Global Classroom Initiative in partnership with the University of Bonn.
DS gratefully acknowledges financial support by the DFG through the funds provided to the Sino--German Collaborative Research Center \mbox{TRR110} ``Symmetries and the Emergence of Structure in QCD'' (DFG Project-ID \mbox{196253076 -- TRR 110}) and by the MKW NRW under the funding code \mbox{NW21-024-A}.  CFR is supported by Spanish Ministerio de Ciencia, Innovación y Universidades \mbox{(MICIU)} under Grant \mbox{No.~BG20/00133}. VM is a Serra Húnter fellow and acknowledges support from the Spanish national Grants PID2023-147112NB-C21 and CNS2022-136085. This work was supported by the U.S. Department of Energy contract \mbox{DE-AC05-06OR23177}, under which Jefferson Science Associates, LLC operates Jefferson Lab, by the U.S. Department of Energy Grant No.~\mbox{DE-FG02-87ER40365}, and contributes to the aims of the U.S. Department of Energy \mbox{ExoHad} Topical Collaboration, contract \mbox{DE-SC0023598}. 

%%%%%%%%%%%%%%%%%%%%%%%%%%%%%%%%%

\appendix
%%%%%%%%%%%%%%%%%%%%%%%%%%%%%%%%%
\section{Hypergeometric formulae} 
\label{app:hypergeometric_formulae}
%%%%%%%%%%%%%%%%%%%%%%%%%%%%%%%%%

We provide a summary of relevant properties of the hypergeometric function and its regularized form. The identities used here are expounded upon in Ref.~\cite{Bateman:1953}. 

The Gaussian hypergeometric function is defined for $|d| < 1$ through its power series:
    \begin{equation}
        \label{eq:gaussF}
        \gaussF{a,b}{c}{d} = \sum_{n=0}^\infty \frac{(a)_n \, (b)_n}{(c)_n} \, \frac{d^n}{n!} ~,
    \end{equation}
where 
    \begin{equation}
        \label{eq:pochhammer}
        (x)_n = \frac{ \Gamma(x+n)}{\Gamma(x)} ~,
    \end{equation}
is the Pochhammer symbol. 
\Cref{eq:gaussF} manifests poles when $c$ is a negative integer and thus we work with the regularized hypergeometric function defined in \cref{eq:gaussFR}, which is pole-free for all values of its arguments. For $|d|<1$ the regularized power series reads:
    \begin{equation}   
        \label{eq:gaussFR_power}
        \gaussFR{a,b}{c}{d} = \sum_{n=0}^\infty \frac{(a)_n \, (b)_n}{\Gamma(c + n)} \, \frac{d^n}{n!} ~.
    \end{equation}
The regularized hypergeometric function is also often represented in an integral form via
    \begin{align}
        \label{eq:reg_hyper_integral}
        \gaussFR{a,b}{c}{d}&\\
        &\hspace{-1.7cm}=\frac{1}{\Gamma(b)\Gamma(c-b)}\int_0^1 \dd x x^{b-1}(1-x)^{c-b-1}(1- xd)^{-a}~,\notag
    \end{align}
which converges only for \mbox{$\Re(c)>\Re(b)>0$} and \mbox{$|\arg(1-d)|<\pi$}.

To consider asymptotic limits, the power series representation of $_2\tilde{F}_1$ in $1/d$ converges as $|d| \to \infty $ and with $a>b$ reads
    \begin{align}
        \gaussFR{a,b}{c}{d}&   \\
        &\hspace{-1.5cm}\to\frac{\Gamma(b-a)}{\Gamma(b)\Gamma(c-a)}(-d)^{-a} + \frac{\Gamma(a-b)}{\Gamma(a)\Gamma(c-b)}(-d)^{-b}~, 
        \notag
    \end{align}
as long as $b-a$ is not an integer.
This latter caveat is important as there are apparent poles emerging from the $\Gamma$ functions in the numerators of both terms in the expansion. As we approach one of these apparent poles, taking $b \to a + n$ for integer $n\geq 1$, the leading term becomes
    \begin{align}
        \gaussFR{a,a+n}{c}{d\to\infty} \to
        \frac{(n-1)!}{\Gamma(a+n) \, \Gamma(c-a)} \, (-d)^{-a}
    \end{align}
without a pole. The specific form of the asymptotic limit that we require is multiplied by additional powers of the parameter $d$ (cf.\ \cref{eq:f(alpha_z)}) and we write this particular case as 
    \begin{align}
        \label{eq:gaussFR_times_d_limit}
        d^j &\, \gaussFR{1+j, j-\alpha}{1+j-\alpha}{d\to\infty}  \to \\
        &\hspace{-0.1cm}(-1)^j\left(\frac{1}{\Gamma(j-\alpha)(-1-\alpha)} \, (-d)^{-1} + \frac{\Gamma(1 + \alpha)}{\Gamma(1+j)} (-d)^{\alpha}\right) ~.
        \nonumber
    \end{align}
Finally, \cref{eq:gaussFR_times_d_limit} seems to manifest a pole in both terms at $\alpha \to -1$, but when taking the limit we see that the result is pole-free:
    \begin{align}
        \label{eq:gaussFR_times_atm1}
        d^j \,\gaussFR{1+j, 1+j}{2+j}{d\to\infty}  & \\
        &\hspace{-1cm}\to\frac{(-1)^j}{j!} \, \log d  \,  (-d)^{-1}~.
        \nonumber
    \end{align}
Although we do not compute the limits for $\alpha \to -j$ for any $j>1$, similar expressions may be derived and \cref{eq:gaussFR_times_d_limit} is finite and pole-free everywhere. 

%%%%%%%%%%%%%%%%%%%%%%%%%%%%%%%%%
\section{Fixed-angle scattering}
\label{app:large_angle}
%%%%%%%%%%%%%%%%%%%%%%%%%%%%%%%%%

While the Regge limit explored in \cref{sec:regge} is of more immediate interest as it relates to the properties of hadrons in different channels, the limit of large $s$ with fixed angle $z_s$, i.e., large $-t$ with the ratio $t/s$ fixed, connects the hadron level amplitude with the inter-meson dynamics of parton exchanges~\cite{Blankenbecler:1973kt,Coon:1974wh}. In particular, while the behavior $s^{\at}$ is characteristic in the Regge limit, at fixed angles, hadronic cross sections exhibit scaling properties of the form
    \begin{equation}
        \label{eq:FA_scaling}
        \frac{\dd\sigma}{\dd t} \simeq \frac{1}{s^2} \, |\Amp(s,t,u)|^2 \propto s^{-N} ~,
    \end{equation}
where the power $N$ is constant and argued to be related to the number of constituent partons involved in the scattering process~\cite{Brodsky:1973kr,Brodsky:1974vy,Matveev:1973ra}.

In this limit, Veneziano-based amplitudes with linear trajectories fall exponentially and are thus entirely unable to reproduce the power-law scaling in \cref{eq:FA_scaling}~\cite{Bugrij:1973ph,Schmidt:1973ew}. More sophisticated dual models such as dual amplitudes with Mandelstam analyticity (DAMA) can be made to exhibit fixed-angle scaling, but at the cost of logarithmically bounded Regge trajectories~\cite{Bugrii:1979zh,Fiore:1999cd}. While a DAMA model of this form can provide a unified description of both fixed-angle and Regge phenomena, the slowly growing RTs make it difficult to also simultaneously describe the resonance spectra at timelike energies. Of note, however, is that DAMA with a logarithmic trajectory allows a microscopic interpretation of its integral representation in terms of meson wave functions and the loop momentum of exchanged partons~\cite{Schmidt:1973ew}.

To investigate under which assumptions the model in \cref{sec:hypergeo_iso} exhibits the scaling behavior, we will use a slightly generalized form of the isobar function with $\jmin = 0$:
    \begin{equation}
        \label{eq:Fas_eta}
        F(\as,\nu_s) = \Gamma(-\as) \, \gaussFR{\eta,-\as}{1 - \as}{\hat{\nu}_s} ~,
    \end{equation}
which reduces to \cref{eq:f(alpha_z)} with $\eta = 1$. 
In the large energy limit at fixed $s$-channel scattering angle $\theta_s$, all three Mandelstam variables are large and we may use
    \begin{equation}
        t \to -s \, \cos^2\frac{\theta_s}{2} \quad \text{and} \quad u \to -s \, \sin^2\frac{\theta_s}{2} ~.
    \end{equation}
We will ignore fixed angular factors and consider only the powers of $s$. This limit thus involves considering $s\to +\infty$ while both $t$ and $u \to -\infty$. With the usual assumptions on the RTs in \cref{sec:regge}, the $s$-channel isobar $\mF{\pm}(s,z_s) \to 0$ exponentially as in \cref{eq:s_limit}. The leading powers of $s$ will thus be given by the $t$- and $u$-channel isobars, which decrease proportional to $t^{-\eta}/\at$ and $u^{-\eta} / \au$, respectively (cf.\ \cref{eq:u_limit}). Since the RTs we consider are unbounded in both directions, as $t \sim u \to -\infty$, both RTs $\at$ and $\au \to -\infty$, and the term of \cref{eq:u_limit} becomes the leading $s$ behavior. Since these will scale with the same powers of $s$, it is sufficient to examine one of these terms. 

Because $\at$ is assumed to satisfy the once-subtracted dispersion relation of \cref{eq:disp_one_sub}, there is an $0 \leq \epsilon < 1$ such that $|\alpha(t\to -\infty)| \lesssim s^{1-\epsilon}$. This means, then, that in the high energy limit at fixed angle, the amplitude \cref{eq:gen_isobar_decomp} will scale as
%5
    \begin{equation}
        \label{eq:amp_scale}
        |\Amp(s,t,u)|^2 \propto \left|\frac{t^{-\eta}}{\alpha(t)}\right|^2 \propto s^{-2(\eta+1-\epsilon)} ~,
    \end{equation}
and the resulting differential cross section as
    \begin{equation}
        \frac{\dd \sigma}{\dd t} \propto s^{-2(\eta+2-\epsilon)} ~.
    \end{equation}
Here we see that regardless of the exact behavior of $\alpha(t)$, so long as it is unbounded in the negative direction, the cross section exhibits fixed power-law scaling behavior with any combination of isobars of the form \cref{eq:Fas_eta}. 
The parameter $\eta$ can be used to change the leading power of $s$ to accommodate theoretical predictions or experimental data. In this way, this isobar model smoothly connects the hadronic component of the isobars, i.e., as encoded through $\as$, to the parton-level dynamics, which are governed by $\eta$. 

In the case of $\pi\pi$ scattering, fixed-angle high-energy scattering is not experimentally possible and we thus do not attempt to fine-tune $\eta$. Instead, we look at the expectation from Regge phenomenology as described in \cref{sec:regge} as well as predictions of hadron scattering processes at large transverse momentum from perturbative QCD (pQCD). Specifically, Refs.~\cite{Blankenbecler:1973kt,Collins:1983fg} argue that because the large $s$- and $|t|$-regime of exclusive hadron processes should be dominated by constituent parton exchanges, all hadron RTs will couple to a lower RT that encodes hard physics and is effectively constant. 
Such a trajectory would correspond to that of a Reggeized \textit{quark} and satisfies $\alpha_q(t\to-\infty) = -1$~\cite{JointPhysicsAnalysisCenter:2024znj}. We thus fix $\eta = - \alpha_q(t) \approx 1$. Identifying the lower bounding behavior of \cref{eq:t_limit_lm1} as arising from a lower-lying quark RT can be trivially considered by generalizing $\eta$ to a function of $t$ but is not considered here. 

With the specific form of \cref{eq:f(alpha_z)} and with $\eta$ fixed as above, the scaling of \cref{eq:FA_scaling} is $N = 6-2\epsilon$. Simple dimensional counting arguments from pQCD predict the elastic scattering of mesons each with two constituents, i.e., a quark and anti-quark pair, have $N = 6$~\cite{Brodsky:1974vy,Matveev:1973ra}. With the trajectories fitted in \cref{sec:application}, the scaling will be dominated by the slowest trajectory, which is that of the $\sigma$. Since our $\alpha_\sigma(s)$ grows logarithmically, it is bounded by any power of $s$ and we can set $\epsilon = 1$ (minus a positive infinitesimal) giving our $\pi\pi$ amplitude a scaling with $N = 4$. As discussed above, the connection of this scaling behavior and pQCD offers an enticing connection between the $\sigma$ and quark exchange dynamics and may offer clues into the internal structure of the former.

%%%%%%%%%%%%%%%%%%%%%%%%%%%%%%%%%
\section{Duality}
\label{app:duality}
%%%%%%%%%%%%%%%%%%%%%%%%%%%%%%%%%

In this appendix, we briefly compare the isobar model constructed in \cref{sec:hypergeo_iso} with commonly held notions of resonance--Regge duality. The concept that $s$-channel poles are ``dual" to $t$-channel poles was first encountered by considering the finite-energy sum rules (FESRs) resulting from the analyticity of scattering amplitudes~\cite{Dolen:1967jr,Chew:1968zz}. If an amplitude is analytic and satisfies crossing symmetry then one may write a dispersion relation up to some finite energy $N$ and derive a self-consistency relation of the form
    \begin{equation}
        \label{eq:FESR}
        \frac{1}{2}\int_{-N}^{N} \dd s^\prime \, \disc \Amp(s^\prime, \, t, \,u) \simeq \sum_i \mathbb{R}_i(N,t) ~.
    \end{equation}
Here, an integral over the discontinuities along both LHC and RHC is related to a sum over Regge terms given by \cref{eq:regge_behavior} on the right-hand side. For simplicity, we have ignored the contributions of the background integral and the possible presence of Regge cuts.

Considering \cref{eq:FESR} in the narrow-width approximation leads to the observation that an amplitude that contains resonances and Regge behavior separately, i.e., $\Amp(s,t,u) = \Amp_\text{Res}(s,t,u) + \Amp_\text{Regge}(s,t,u)$, leads to violations of the FESRs (and therefore analyticity) in intermediate energies where the two terms can interfere, i.e., the ``double counting" problem~\cite{Dolen:1967jr,Schmid:1968zz}. A proposed solution was that the resonance and Regge behaviors are dual to each other, in the Dolen--Horn--Schmid (DHS) sense, meaning they must arise from a single function when evaluated in different limits and there is therefore no second term with which to interfere. This led to the general lore that amplitudes will fall into one of two distinct categories: dual models, which realize this DHS duality, or interference models, which sum direct and crossed channel processes like in a Feynman-diagram-based approach. Since only the former is thought to satisfy the FESRs, it is believed to be more fundamental. 

The isobar model constructed in \cref{sec:hypergeo_iso} belongs in the latter category with explicit sums of poles in the $s$-, $t$-, and $u$-channels. The model in \cref{eq:gen_isobar_decomp}, however, also satisfies the FESRs by construction as evident by the analyticity of \cref{eq:f(alpha_z)}, the cut structure in \cref{fig:cut_structure_fullamplitude}, and the Regge asymptotics in \cref{sec:regge}. Indeed, whether satisfying the FESRs requires amplitudes to be dual in the DHS sense has been criticized in Refs.~\cite{Alessandrini:1968hyk,Barger:1968sxq,Alessandrini:1969hx,Childers:1969am,Kellett:1970yx}, since interference models can be constructed to satisfy \cref{eq:FESR} when going beyond the narrow-resonance approximation and introducing cuts. Our model is one realization of these ideas, with the cuts of the RTs playing a pivotal role in the Regge asymptotics (cf.\ \cref{eq:s_limit}).

Other works further criticized the distinction between dual and interference with proofs that, under general assumptions, amplitudes can always be decomposed into separate resonance and Regge terms akin to an interference model~\cite{Jengo:1969fb,Lichtenberg:1969cm,Venturi:1970eu}. 
In light of this seemingly blurry boundary between the two classes of models, Ref.~\cite{Childers:1969am} suggests an unambiguous statement of duality is that resonances and Regge behavior are dual if they both arise from the same function $\alpha(s)$ with respect to poles in the complex $j$-plane. As explored in \cref{sec:regge,sec:resonances}, this is the case for our model.

In addition to the general considerations above, we can explicitly compare the implementation of our model to $\pi\pi$ scattering, discussed in \cref{sec:application} and expressed in the charge basis in \cref{app:isospin}, to dual amplitudes for the same reaction, specifically the Veneziano--Lovelace--Shapiro (VLS) model~\cite{Veneziano:1968yb,Lovelace:1968kjy,Shapiro:1969km} and the broader class of dual amplitudes with Mandelstam analyticity (DAMA) models (for a general survey of the latter, see Refs.~\cite{Cohen-Tannoudji:1971mlv,Bugrij:1972ce,Bugrij:1973ph} and references therein).   
 
In a generic dual model, the charge basis $\pi\pi$ amplitude of \cref{eq:chargebasis} is decomposed in terms of a single function of two variables~\cite{Shapiro:1969km}:
    \begin{equation}
        \label{eq:M_LS}
        \M(s,t,u) = V(s,t) + V(s,u) - V(t,u) ~,
    \end{equation}
where $V(s,t) = V(t,s)$ contains resonances in either the $s$- or $t$-channel in different limits. The structure of \cref{eq:M_LS} is fairly rigid to ensure the absence of $I=2$ resonances when constructing isospin amplitudes with \cref{eq:AI_from_M}~\cite{Sivers:1971ig}. The choice of function $V(s,t)$, however, can be shown to be fairly general and, in the language of DAMA, typically defined in terms of an integral of the form
    \begin{equation}
        \label{eq:V_DAMA}
        V(s,t) = \int_0^1 dx \, x^{-\as-1} \, (1-x)^{-\at-1} \, g(s,t, x) ~,
    \end{equation}
with a kernel function $g(s,t,x) = g(t,s,1-x)$, which is regular at the end points of integration. A classical choice of the kernel is a constant, i.e., $g(s,t,x) = g^2$, which reproduces the Euler beta function:
    \begin{equation}
        \label{eq:V_venez}
        V(s,t) = g^2 \, \frac{\Gamma(-\as) \, \Gamma(-\at)}{\Gamma(-\as-\at)} ~,
    \end{equation}
and is the basis of the VLS model. More complicated functions have also been considered, cf.\ Ref.~\cite{Suzuki:1969zq}. The integral in \cref{eq:V_DAMA} does not typically converge for all $s$ and $t$ and must be extended, e.g., to the resonance region $\as >0$, through analytic continuation~\cite{Bugrij:1973ph}. 

We want to consider whether the isobar model constructed in \cref{sec:application} can recover the structure of \cref{eq:M_LS,eq:V_DAMA}, despite not appearing as a dual model. First, the absence of $I=2$ is trivially achieved in \cref{eq:mA_I} by decoupling any $s$-channel RTs in that channel.  Thus, we set all couplings in $I=2$ to zero such that \cref{eq:chargebasis} reduces to:
    \begin{equation}
        \label{eq:M_simp}
        \M(s,t,u) = \frac{1}{3} \mF{0}(s,z_s) + \frac{1}{2} \left[\mF{1}(t,z_t) - \mF{1}(u,z_u) \right] ~.
    \end{equation}
The remaining $I=0$ and $1$ isobars can then be assumed to be given by a single exchange degenerate trajectory $\alpha(s)$ with $\jmin = 0$ and the couplings in the different isospins related to a universal coupling by $g^2 = g^2_0/3 = g_1^2/2$ (cf.\ the exchange degeneracy assumptions in \cref{eq:EXD}). 
With \cref{eq:mF_I} and these choices, \cref{eq:M_simp} can be cast in the same form as \cref{eq:M_LS} using:
    \begin{align} 
        \label{eq:V_ours}
        V(s,t) &= \frac{g^2}{2} \left[F(\as, \, \nu_s)  + F(\at, \,\nu_t)\right] ~, \nonumber \\ 
        V(s,u) &= \frac{g^2}{2} \left[F(\as, \, -\nu_s) + F(\au, \, -\nu_u)\right] ~, \\
        V(t,u) &= \frac{g^2}{2} \left[F(\at, \, -\nu_t) + F(\au, \, \nu_u)\right] ~, \nonumber
    \end{align}
in terms of the isobar functions in \cref{eq:f(alpha_z)}. 
These functions are symmetric in their arguments and related only by the interchange of Mandelstam variables. As discussed in \cref{sec:resonances}, near a resonance pole in the complex plane, $V(s \to s_j,t) \propto P_j(z_s)/(j-\as)$, while simultaneously $V(s, t\to t_j) \propto P_j(z_t)/(j-\at)$. Further, the spectrum of the Chew--Frautschi plot of \cref{eq:V_ours} matches that of the \cref{eq:V_venez} (albeit not on linear, real trajectories), since $V(s,t)$ is not yet symmetrized with a definite parity. 
When evaluating the Regge limits with the usual assumptions in \cref{sec:regge}, we have $V(s\to\infty, t ) \propto s^{\at}$ and $V(s, t \to \infty) \propto t^{\as}$, once again as in \cref{eq:V_venez}.

Unlike \cref{eq:V_venez}, however, these limits are achieved through the sum of infinitely many terms in both $(j-\as)^{-1}$ and $(j-\at)^{-1}$. Thus, a limiting case of our model can recover all the same phenomenological features of the VLS amplitude, but without DHS duality. Moreover, our amplitude easily loosens the assumptions of exchange degeneracy, zero-width resonances, and linear trajectories, which have been notorious difficult challenges for Veneziano-like amplitudes to overcome.  

We can also compare \cref{eq:V_ours} to the more general class of DAMA models given by \cref{eq:V_DAMA}. Using the integral form of the regularized hypergeometric function, \cref{eq:reg_hyper_integral}, we can write the isobar in \cref{eq:f(alpha_z)} as:
    \begin{equation}
        \label{eq:F_int}
        F(\as, \, \nu_s) = \int_0^1 \dd x \, \frac{x^{-\as -1}}{1-\hat{\nu}_s \, x} ~,
    \end{equation}
which converges so long as $\Re\as < 0$.
With this representation, \cref{eq:V_ours} is of the form \cref{eq:V_DAMA} with the choice
    \begin{equation}
        \label{eq:g_ours}
        g(s,t,x) = \frac{g^2}{2} \left[ \frac{(1-x)^{\at+1}}{1-\hat{\nu}_s \, x} +  \frac{x^{\as+1}}{1-\hat{\nu}_t \, (1-x)} \right] ~, 
    \end{equation}
which satisfies the necessary symmetries and is regular at $x = 0$ and 1. This seems to suggest a close connection between our amplitude and DAMA, albeit with a choice of a kernel that separates $s$- and $t$-channel components and thus breaks the intrinsic duality. Such a splitting of DAMA into an interference-like representation has been considered in Refs.~\cite{Venturi:1970eu,Bugrij:1973ph}, although with different model constructions. \Cref{eq:g_ours} is a particularly convenient choice as it allows the analytic continuation to arbitrary $\as$ and $\at$ to be done in closed form via the hypergeometric function in \cref{eq:f(alpha_z)}. 

Because we are able to write the isobars in the form of the DAMA integral and have scaling properties as explored in \cref{app:large_angle}, we conjecture a similar microscopic interpretation of the kernel in \cref{eq:g_ours} in terms of meson wave functions and interchanged parton momentum as in Ref.~\cite{Blankenbecler:1973kt} but do not explore this here.

%%%%%%%%%%%%%%%%%%%%%%%%%%%%%%%%%
\section{Isospin structure} 
\label{app:isospin}
%%%%%%%%%%%%%%%%%%%%%%%%%%%%%%%%%

Here, we show that \cref{eq:mA_I} is indeed crossing symmetric by computing the amplitudes projected onto the $t$- and $u$-channel isospin bases. The $s$--$t$ isospin crossing matrix is already defined in \cref{eq:Cmatrix} and we define two more matrices~\cite{Ananthanarayan:2000ht}
\begin{subequations}
    \begin{equation}
        \label{eq:C_su}
        C_{su} =
        \setlength\arraycolsep{4pt}
        \def\arraystretch{1.5}
        \begin{pmatrix}
        \frac{1}{3}& -1&\frac{5}{3}\\
        -\frac{1}{3}&\frac{1}{2}&\frac{5}{6}\\
        \frac{1}{3}&\frac{1}{2}&\frac{1}{6}\\
        \end{pmatrix}
    \end{equation}
and 
    \begin{equation}
        C_{tu} =
        \setlength\arraycolsep{4pt}
        \def\arraystretch{1.5}
        \begin{pmatrix}
        1 & 0 & 0 \\
        0 & -1 & 0 \\
        0 & 0 & 1 \\
        \end{pmatrix}~.
    \end{equation}
\end{subequations}
We see the elements of \cref{eq:C_su} are related to those in \cref{eq:Cmatrix} by $C_{st}^{I\Ip} = (-1)^{I+\Ip} \, C_{su}^{I\Ip}$. Each matrix obeys $C_{st}^2 = C_{su}^2 = C_{tu}^2 = 1$ as well as the cyclic relations
    \begin{subequations}
        \begin{equation}
            C_{st} \, C_{tu} = C_{tu} \, C_{su} =  C_{su} \, C_{st}
        \end{equation}
        and
        \begin{equation}
            C_{su} \, C_{tu} = C_{tu} \, C_{st} = C_{st} \, C_{su} ~. 
        \end{equation}
    \end{subequations}

These matrices arise as coefficients when considering isospin amplitudes in different frames. For example, defining a vector with respect to isospin components, $\vec{\Amp}(s,t,u) = \{\Amp^0(s,t,u), \, \Amp^1(s,t,u), \, \Amp^2(s,t,u)\}$, crossing symmetry requires the $t$-channel isospin amplitudes to fulfill
\begin{subequations}
    \label{eq:Iso_crossing}
    \begin{equation}
        \vec{\Amp}(t,s,u) = C_{st} \, \vec{\Amp}(s,t,u) ~,
    \end{equation}
and similarly for the $u$-channel:
    \begin{equation}
         \vec{\Amp}(u,t,s) = C_{su} \, \vec{\Amp}(s,t,u) ~.
    \end{equation}
\end{subequations}

Defining an analogous vector for the isospin-definite isobars in \cref{eq:mF_I}, $\vec{\F}(s,z_s) = \{\mF{0}(s,z_s), \, \mF{1}(s,z_s), \, \mF{2}(s,z_s)\}$, we can write \cref{eq:mA_I} compactly as:
    \begin{equation}
        \label{eq:A1_vec}
        \vec{\Amp}(s,t,u) = \vec{\F}(s,z_s) + C_{st} \, \vec{\F}(t,z_t) + C_{su} \, \vec{\F}(u, z_u) ~.
    \end{equation}
Then, we calculate
\begin{subequations}
    \begin{align}
        C_{st} \, &\vec{\Amp}(s,t,u) \nonumber \\
        &= C_{st} \, \vec{\F}(s,z_s) + \vec{\F}(t,z_t) +  C_{su} \, C_{tu} \,\vec{\F}(u, z_u) \nonumber \\
        &= C_{st} \, \vec{\F}(s,z_s) + \vec{\F}(t,z_t) +  C_{su} \, \vec{\F}(u,-z_u) \nonumber \\
        & = \vec{\Amp}(t,s,u)~,
    \end{align}
where $C_{tu} \, \vec{\F}(x,z_x) = \vec{\F}(x,-z_x)$ is used. For the $u$-channel, the calculation proceeds identically:
    \begin{align}
        C_{su} \, &\vec{\Amp}(s,t,u) \nonumber \\
        &= C_{su} \, \vec{\F}(s,z_s) + C_{st} \, \vec{\F}(t,-z_t) + \vec{\F}(u, z_u) \nonumber \\
        & = \vec{\Amp}(u,t,s)~,
    \end{align}
\end{subequations}
and thus \cref{eq:A1_vec} indeed satisfies \cref{eq:Iso_crossing}. 

The starting point of our construction \cref{eq:mA_I} are isobars that already carry definite isospin. However, we may also compare to the charge basis, which is the traditional starting point of narrow-resonance models, such as in Refs.~\cite{Lovelace:1968kjy,Shapiro:1969km}. By defining the generic isospin amplitude
    \begin{align}
        &\mathcal{M}^{abcd}(s,t,u) = \\
        &~ \delta
        ^{ab}\delta^{cd} \, \M(s,t,u) + 
        \delta^{ac}\delta^{bd} \, \M(t,s,u) + \delta^{ad}\delta^{bc} \, \M(u,t,s) ~,  \nonumber
    \end{align}
with isospin indices $a$, $b$, $c$, and $d$ in the Cartesian basis, the amplitude $\M(s,t,u)$ can be related to \cref{eq:mA_I} by
    \begin{equation}
        \label{eq:AI_from_M}
        \begin{pmatrix}
            \Amp^0(s,t,u) \\
            \Amp^1(s,t,u) \\
            \Amp^2(s,t,u) 
        \end{pmatrix} = 
        \begin{pmatrix}
            3 & 1 & 1 \\
            0 & 1 & -1 \\
            0 & 1 & 1 
        \end{pmatrix}
        \begin{pmatrix}
            \M(s,t,u) \\
            \M(t,s,u) \\
            \M(u, t, s) 
        \end{pmatrix}
        ~.
    \end{equation}
This relation may be inverted yielding the charge basis amplitude in terms of isobars:
    \begin{align}
        \label{eq:chargebasis}
       &\M(s,t,u) = \frac{1}{3} \, \mF{0}(s,z_s) + \frac{1}{2} \left[\mF{1}(t,z_t) - \mF{1}(u,z_u) \right] \nonumber \\
        &- \frac{1}{3} \left[\mF{2}(s,z_s) - \frac{3}{2}\left(\mF{2}(t,z_t) + \mF{2}(u,z_u) \right) \right]  ~.
    \end{align}
%%

%%%%%%%%%%%%%%%%%%%%%%%%%%%%%%%%%
\section{Pole positions}
\label{app:poles}
%%%%%%%%%%%%%%%%%%%%%%%%%%%%%%%%%

In this appendix, we describe the procedure to extract the pole positions using the RTs constructed in the text. The resonances of a given isobar \cref{eq:f(alpha_z)} correspond to integer values of $\as$ in the complex $s$ plane. Thus determining the location for a resonance with spin-$j$ on a trajectory $\as$ entails searching for roots of $(j-\alpha(s))$ in the lower-half complex plane. Because the trajectories are defined through \cref{eq:disp_one_sub}, they are evaluated on the first Riemann sheet and must be continued to the unphysical sheet. This can be accomplished in a number of ways, but we use the Schlessinger Point Method (SPM)~\cite{Schlessinger:1968vsk,Tripolt:2016cya,Binosi:2019ecz,Binosi:2022ydc,Pelaez:2022qby}. From $\alpha(s)$ evaluated through \cref{eq:disp_one_sub}, we construct the continued fraction:
    \begin{equation}
    \label{eq:continued_fraction}
    \tilde{\alpha}_N(s)=\alpha\left(s_{1}\right)\Big/
    \Bigg(1+\frac{a_{1}\left(s-s_{1}\right)}{1+\frac{a_{2}\left(s-s_{2}\right)}{\ddots a_{N}\left(s-s_{N}\right)}}\Bigg) ~,
    \end{equation}
where the coefficients $a_k$ for $k = 1, \dots, N$ are chosen such that $\tilde{\alpha}(s_k) = \alpha(s_k)$ at $N+1$ energy points on the real line $s_k$. We refer to the appendix of Ref.~\cite{Schlessinger:1968vsk} for the explicit formula to calculate each $a_k$ recursively. If the points $s_k$ interpolate a segment above threshold, then the resulting $\tilde{\alpha}(s)$ can be used to probe the unphysical Riemann sheet in the lower-half plane, which will be smoothly connected to the first Riemann sheet in the upper-half plane.

The interpolated $s_k$'s are chosen to be evenly spaced points from threshold to some $s_\text{max}$. For a given $N$ and $s_\text{max}$, then, poles are found by minimizing $|j-\tilde{\alpha}(s)|$ for the desired value of $j$. Once poles are found, their stability upon varying $N$ and $s_\text{max}$ must be verified as an analytic continuation using the SPM may introduce spurious singularities.

The above procedure can be done with the trajectory alone. In order to also extract the residue of the poles, however, we need to use the RT in conjunction with the fitted isobar model.
Since the pole at $\alpha(s_j) = j$ will appear in the PW projection of the direct channel isobar $f_j^I(s)$ in \cref{eq:f}, the residue can be extracted by evaluating
    \begin{equation}
        \label{eq:residue}
        g^2 = - 16\pi \lim_{s\to s_j}(s-s_j)  \, \frac{(2j+1)}{(2  q_s)^{2j}} \, f_j^I(s)~.
    \end{equation}
The PW projected isobar must also be continued to the contiguous Riemann sheet, which can be accomplished using the SPM as above, or directly from \cref{eq:f} by replacing $\alpha(s)$ with \cref{eq:continued_fraction}. Once again, the stability of the residue must be checked by varying $N$ and $s_\text{max}$.

Because the RTs naturally sort resonances into different particle families each entering through a separate $F(\alpha_i(s), \, \nu_s)$ in \cref{eq:mF_pm}, as we approach the pole, the residue in \cref{eq:residue} will actually only depend on a single term in the sum of \cref{eq:f}. This means the extraction of pole parameters in situations involving multiple resonances appearing on different RTs can be simplified, since, after the amplitude is fit to data, one only needs to examine a single trajectory at a time.

\bibliographystyle{apsrev4-1} 
\bibliography{Literature}

\end{document}